\documentclass[a4paper,11pt]{article}
\pdfoutput=1 

\usepackage{jcappub} 

\usepackage[T1]{fontenc} 

\usepackage{natbib}

\usepackage{graphicx}	
\usepackage{amsmath}	
\usepackage{amssymb}	
\usepackage{float}

\usepackage{dcolumn}
\usepackage{hhline}
\usepackage{booktabs}

\usepackage{bm}
\usepackage{bbold}
\usepackage{comment}  

\newif\ifshow
\showfalse  

\ifshow
  \includecomment{hide}
\else
  \excludecomment{hide}
\fi

\DeclareMathOperator{\cov}{Cov}

\newcommand{\vk}{{\bm k}}
\newcommand{\vq}{{\bm q}}
\newcommand{\vx}{{\bm x}}
\newcommand{\vs}{{\bm s}}

\newcommand{\nbar}{\bar{n}}
\newcommand{\W}{\mathcal{W}}
\renewcommand{\S}{\mathcal{S}}
\newcommand{\Q}{\mathcal{Q}}

\newcommand{\QW}[1][]{\Q_{\W #1}}
\newcommand{\QS}[1][]{\Q_{\S #1}}
\newcommand{\QX}[1][]{\Q_{\times #1}}

\newcommand{\disc}{\mathrm{disc}}
\newcommand{\conn}{\mathrm{conn}}

\newcommand{\ensavg}[1]{\bigl\langle #1 \bigr\rangle}

\title{RSD measurements from BOSS galaxy power spectrum using the halo perturbation theory model}

\author[a,b]{Byeonghee Yu}
\author[a,b,c]{Uro{\v s} Seljak}
\author[d,e]{Yin Li}
\author[f]{Sukhdeep Singh}

\affiliation[a]{Department of Physics, University of California, Berkeley, CA 94720, USA}
\affiliation[b]{Berkeley Center for Cosmological Physics, University of California, Berkeley, CA 94720, USA}
\affiliation[c]{Lawrence Berkeley National Laboratory, One Cyclotron Road, Berkeley, CA 94720, USA}
\affiliation[d]{Department of Mathematics and Theory, PengCheng Laboratory, Shenzhen, Guangdong 518066, China}
\affiliation[e]{Center for Computational Astrophysics \& Center for Computational Mathematics, Flatiron Institute, New York, NY 10010, USA}
\affiliation[f]{McWilliams Center for Cosmology, Carnegie Mellon University, Pittsburgh, PA 15213, USA}

\emailAdd{bhyu@berkeley.edu}
\emailAdd{useljak@berkeley.edu}
\emailAdd{eelregit@gmail.com}
\emailAdd{sukhdeep@cmu.edu}

\abstract{We present growth of structure constraints from the cosmological analysis of the power spectrum multipoles of SDSS-III BOSS DR12 galaxies. We use the galaxy power spectrum model of
\cite{Hand:2017ilm}, which
decomposes the galaxies into
halo mass bins, each of
which is modeled separately using the relations between halo biases and halo mass. The model combines Eulerian perturbation theory and halo model
calibrated on $N$-body simulations to model the halo clustering. In this work, we also generate the covariance matrix by combining the analytic disconnected part with the empirical connected part: we smooth the connected component by selecting a few principal components and show that it achieves good agreement with the mock covariance. 
Our analysis differs from 
recent analyses in that we 
constrain a single parameter
$f\sigma_8$ fixing everything 
else to Planck+BAO prior, thereby
reducing the effects of prior 
volume and mismodeling. 
We find tight constraints on $f\sigma_8$: $f\sigma_8(z_{\mathrm{eff}}=0.38)=0.489 \pm 0.038$ and $f\sigma_8(z_{\mathrm{eff}}=0.61)=0.455 \pm 0.028$ at $k_{\mathrm{max}} = 0.2\ h$Mpc$^{-1}$, with an overall amplitude error of 5\%, and
in good agreement 
(within 0.3 sigma) of Planck amplitude. 
We discuss the sensitivity of cosmological parameter estimation to the choice of scale cuts, covariance matrix, and the inclusion of hexadecapole $P_4(k)$. We show that with $k_{\mathrm{max}} = 0.4\ h$Mpc$^{-1}$ the constraints
improve considerably to an overall 3.2\% amplitude error,
but there is some evidence of model misspecification on MultiDark-PATCHY mocks. Choosing $k_{\mathrm{max}}$
consistently and
reliably remains the main
challenge of RSD analysis methods.
}

\begin{document}
\maketitle
\flushbottom

\section{Introduction}

Large-scale clustering of the galaxies in redshift surveys is one of the major cosmological probes which gives us insight into gravity, dark energy, and primordial non-Gaussianities. We can quantify this structure using the 2-point correlation function or the power spectrum.
The 2-point analyses have made accurate measurements of baryon acoustic
oscillations (BAO), caused by sound waves in the pre-decoupling Universe
\citep{EisensteinHu98}. The BAO data have both isotropic and anisotropic
components, and with galaxy samples from Baryon Oscillation
Spectroscopic Survey (BOSS), a part of Sloan Digital Sky Survey
(SDSS)-III, it provides constraints on the distance scale with a
percent-level precision \cite{AlamEtAl17}.

Galaxy clustering amplitude
cannot be directly related
to the dark matter amplitude
due to galaxy biasing. However,
we can consider
another kind of anisotropy in the clustering of galaxies caused by the redshift-space distortions (RSD). It is created by peculiar velocities of galaxies, affecting the measured clustering signal in redshift space along the line-of-sight, but not transverse
to it. Such distortions depend on the underlying matter density field, which are correlated with the velocity field. In the linear regime we parametrize RSD with the parameter $\beta = f/b$, where $f$ is the linear growth rate, and $b$ is the galaxy bias \citep{Kaiser87}. On small scales the linear theory breaks down, and non-linear distortions, such as the Finger-of-God (FoG) effect, clustering dilution in redshift space along the line of sight due to the motion of galaxies within virialized dark matter halos, need to be accounted for.

RSD has become one of the most powerful cosmological probes by measuring the growth of structure via the parameter combination $f(z)\sigma_8(z)$, thereby testing dark energy and different gravity models. \cite{Reid:2014iaa} provides the 2.5\% constraint on $f\sigma_8$ on the BOSS CMASS galaxies using a simulation-based analysis, but \cite{Reid:2014iaa} does not employ an analytic approach to model the small-scale clustering and uses only a single simulation box. \citep{AlamEtAl17} presents the Data Release 12 (DR12) final consensus results on the BOSS galaxies, over the redshift range $0.2 < z < 0.75$, and provides 9.3 and 8.0\% $f\sigma_8$ constraints on low-redshift ($z_{\mathrm{eff}}=0.38$) and high-redshift galaxies ($z_{\mathrm{eff}}=0.61$). More recently, there are BOSS DR12 measurements of the growth of structure from PT-based models \citep{Ivanov:2019pdj, DAmico:2019fhj, Zhang:2021yna, Chen:2021wdi} and from simulation-based models \citep{Lange:2021zre, Kobayashi:2021oud, Zhai:2022yyk, Yuan:2022jqf}. In particular, \citep{Zhai:2022yyk} proposes a Gaussian Process emulator and provides 7.4, 5.6, and 4.9\% constraints on $f\sigma_8$ at $z_{\mathrm{eff}}=0.25, 0.4$ and $0.55$, and \citep{Yuan:2022jqf} develops a hybrid emulator which combines emulator with Markov chain Monte Carlo (MCMC) sampling, giving the 3.6\% constraint on the BOSS CMASS galaxies. In addition, some of the recent works measure the clustering of the DR16 extended BOSS (eBOSS) samples and provides the growth of structure measurements \citep{Ivanov:2021zmi, Chapman:2021hqe}.

This work applies the redshift-space galaxy power spectrum model of \citep{Hand:2017ilm} to the BOSS DR12 galaxy samples. \citep{Hand:2017ilm} proposes an approach which combines perturbation theory(PT)-based modeling techniques and simulation-based analyses. Following the halo model formalism in \cite{Okumura:2015fga}, this model decomposes a galaxy sample into centrals and satellites and separately model the 1-halo and 2-halo---correlations of 2 galaxies in the same and different halos, respectively---contributions to the clustering of galaxies. The dark matter halo power spectrum model in redshift space is based on the distribution function approach \citep{Seljak:2011tx, Okumura12JCAP, 2012JCAP...02..010O, 2012JCAP...11..009V, Vlah:2013lia, Blazek2011}, and Eulerian PT and halo biasing model are used to model the underlying dark matter correlator terms \citep{Vlah:2013lia}. Then, some of the key terms in the models are calibrated from the results of N-body simulations. \citep{Hand:2017ilm} tests and validates this power spectrum model by performing independent tests using high-fidelity, periodic $N$-body simulations and realistic BOSS CMASS mocks, showing that the recovered values of $f\sigma_8$ has only small bias.

In this work, we not only extend the work from \citep{Hand:2017ilm} by applying its power spectrum model to the BOSS DR12 galaxies but also develop the hybrid covariance matrix, which combines the analytic disconnected part \citep{Li2019} and empirical connected part, including up to four principal components. Such covariance matrix can be especially useful for the analysis of next-generation redshift surveys. We also discuss how the choice of scale cuts, covariance matrix, and the inclusion of hexadecapole $P_4(k)$ affect our clustering analysis and show that removing BAO information from the multipole measurements only affects cosmological parameter estimation in a negligible way. We also compare our growth of structure constraints with other BOSS DR12 measurements in the literature, based on both PT-based models and simulation-based models.

In this paper 
we adopt a different 
approach than many of the 
recent papers 
\citep{AlamEtAl17,Ivanov:2019pdj,Chen:2021wdi,Kobayashi:2021oud,DAmico:2019fhj,Yuan:2022jqf,Lange:2021zre,Zhai:2022yyk}
in that 
we only fit a single parameter 
$f\sigma_8$ to the RSD data, 
fixing everything else. There 
are several reasons for this 
approach. First of all, 
angular clustering 
in linear regime of redshift space is 
directly probing $f\sigma_8$, 
and it is reasonable to 
analyze the parameter that 
is closest to the data. 
Many of recent analyses 
of BOSS data find a low 
amplitude compared to 
Planck (see e.g. Fig. \ref{fig:BOSS_con}), but the comparison 
is not straight-forward since 
they typically fit several 
parameters at once. 
Fitting several parameters to RSD data raises several issues: as the 
data are not sensitive to all 
of the parameters the priors 
become important, and these 
may bias the posteriors of 
parameters the data are 
sensitive to. 

Second issue 
has to do with model 
misspecification. All of 
the RSD models are likely 
misspecified to some extent: 
there are significant biases
in parameter fits 
already at 
$k=0.2$h/Mpc \cite{2020Nishimichi}, which 
is the scale typically adopted in the 
data analysis, so there is 
model misspecification already at that 
scale. Since RSD is strongly 
correlated with AP parameter 
\cite{2017Sanchez} $F_{AP}(z)=D_A(z)H(z)(1+z)$, where $D_A$ is 
angular diameter distance, 
$H(z)$ is Hubble parameter 
and $z$ is the redshift, this 
problem can be exacerbated, 
since even a small model
misspecification can lead 
to a large bias in the 
parameter fits due to this correlation: this is seen in the fits to the model used in this paper \cite{Hand:2017ilm}, where 
the bias of free AP parameter can be larger than the bias 
of fixed AP parameter. For 
this reason fixing AP 
parameter may be less prone to 
bias in the parameter fits. 
Furthermore, AP parameter is 
well measured by Planck+BAO (or by BBN+BAO). Indeed, 
$\lim_{z=0}[F_{AP}/z]\equiv 1$ by definition, so 
at low redshifts there is 
no point in fitting for $F_{AP}$ at all. For CMASS and Lowz data used in this paper 
the uncertainty in $F_{AP}$ from Planck+BAO \cite{Planck:2018vyg} is 
very small compared to the 
precision of AP parameter 
we can achieve from RSD: $F_{AP}(z=0.38)=0.584\pm 0.001$  and $F_{AP}(z=0.61)=1.18\pm 0.003$. For 
this reason we fix $F_{AP}$
to Planck+BAO value. We have 
verified that varying $F_{AP}$
within Planck+BAO uncertainties makes no effect on the derived $f\sigma_8$
value. 

A third 
issue is that we want to 
compare RSD analysis to Planck, but if we fit several 
parameters we must compare all 
of them, and account for 
multiple comparisons by 
the appropriate Look Elsewhere 
factor. However, RSD is really 
sensitive to $f\sigma_8$ parameter
only, which argues for doing 
a single parameter analysis and compare its value to Planck. Our primary motivation 
is to compare BOSS RSD analysis
to Planck, rather than to 
other RSD analyses which 
have adopted a different 
methodology. We want to 
phrase the comparison as 
one of (dis)agreement between 
Planck and BOSS RSD, and 
for this reason we simply 
fix everything except the 
parameter that BOSS RSD is 
most sensitive to. 

The remainder of this paper is organized as follows. In section~\ref{2}, we describe the mock simulations used to validate our model and the actual galaxy sample from BOSS DR12 used for the main analysis. Section~\ref{3} presents the galaxy power spectrum estimator and model parameters, as well as the survey window function convolved with the model. In section~\ref{4}, we outline analysis methods for the cosmological parameter estimation and introduce the hybrid covariance matrix, demonstrating its accuracy compared to the mock covariance matrix. In section~\ref{6}, we validate the model performance by providing test results on the mock catalogues which mimic the BOSS DR12 target selection. In section~\ref{7}, we discuss the main results of this paper and conclude in section~\ref{8}.

\section{Data}\label{2}
\subsection{SDSS-III BOSS DR12}
In this work, we use the spectroscopic galaxy samples from SDSS-III BOSS DR12 \citep{Blanton:2003,Bolton:2012,Ahn:2012,Dawson:2013,Smee:2013,Alam2015}, selected using the imaging data from earlier SDSS-I and SDSS-II surveys. The BOSS DR12 samples are divided into the three redshift bins - z1 ($0.2<z<0.5$), z2 ($0.4<z<0.6$), and and z3 ($0.5<z<0.75$), following \cite{AlamEtAl17}. Because z2 overlaps with the other two samples and thus gives results correlated with others, we only consider z1 and z3, two non-overlapping BOSS DR12 samples. Each sample is observed in two different patches on the sky: North Galactic Cap (NGC) and South Galactic Cap (SGC). The BOSS DR12 sample, covering the redshift range $0.2<z<0.75$ over the area of 10,252 deg$^2$, contains 1,198,006 galaxies - 864,924 in NGC and 333,082 in SGC, and due to the difference in the imaging of the NGC and SGC samples, they have different characteristics, such as the bias parameter, and therefore we run an independent analysis on each sky region.


To address the problems arising from incompleteness of the BOSS survey, we apply weights to the galaxies, where the weights are
given by
\begin{equation}
	w=w_{\mathrm{sys}}(w_{\mathrm{no-z}}+w_{\mathrm{cp}}-1),
\end{equation}
where $w_{\mathrm{sys}}$ is a systematic weight. $w_{no-z}$ and $w_{\mathrm{cp}}$ correct for missing redshifts due to
failure to obtain redshift (no-z) and fiber collisions for close pairs (cp) \citep{Ross2012}.

The effective redshift for the z1 and z3 samples can be obtained as
\begin{equation}
    z_{\mathrm{eff}} = \frac{\sum_i^{N_{\mathrm{gal}}} w_{\mathrm{fkp},i} \cdot w_{i} \cdot z_i}{\sum_i^{N_{\mathrm{gal}}} w_{\mathrm{fkp},i} \cdot w_{i}},
\end{equation}
where $w_{\mathrm{fkp}} = (1+\bar{n}(z)P_0)^{-1}$ with $P_0 = 10^4 h^{-3}$Mpc$^3$.
We find that $z_{\mathrm{eff}} = 0.38$ and 0.61, respectively for z1 and z3.

\subsection{MultiDark-PATCHY mock catalogues}\label{2.2}

We use the MultiDark(MD)-PATCHY mock catalogues \citep{KitauraEtAl16} for the BOSS DR12 dataset, produced using approximate gravity solvers and galaxy biasing models calibrated to the BigMultiDark simulations, which use $3840^3$ particles on a volume of $(2.5h^{-1}Mpc)^3$, and it reproduces the observed evolution of the clustering of the BOSS DR12. All quantities in these catalogues assume Planck13 cosmology: $\Omega_m = 0.307115, \Omega_L = 0.692885, \Omega_b = 0.048, \sigma_8 = 0.8288$ and $h = 0.6777$. We have 2048 mock catalogues available for both NGC and SGC hemispheres.

In section~\ref{4.1}, we use Version 6C (V6C) catalogues, which is adjusted to reproduce the observed clustering measurements of the BOSS DR12, to evaluate the mock covariance matrix. Figure~\ref{fig:WF2} shows that the power spectrum multipoles of V6C catalogues (colored dotted curves) match well with the BOSS DR12 measurements (circular data points). In section~\ref{3.1}, we use Version 6S (V6S) catalogues to confirm that our theoretical model is accurate enough to obtain the cosmological parameter constraints. However, the difference between V6C and V6S catalogues are only subtle \cite{Beutler:2018vpe}.

\section{Redshift-space galaxy power spectrum}\label{3}
\subsection{Model: perturbation theory}\label{3.1}

In this work, we use the galaxy power spectrum model of \cite{Hand:2017ilm}, which is based on perturbation theory combined with simulation-based calibration of halo model terms. We only briefly summarize the model here and refer the reader to \cite{Hand:2017ilm} for more details.

This model follows the halo model formalism in \cite{Okumura:2015fga}, separately modeling the 1-halo and 2-halo contributions to the correlation of central and satellite galaxies. For this modeling, we decompose the galaxy sample into four sub-samples, based on whether there exists at least one other neighboring satellite in the given halo: isolated centrals without satellites (``type A'' centrals), centrals with one or more satellites (``type B'' centrals), isolated satellites (``type A'' satellites), and non-isolated satellites (``type B'' satellites). Such sub-sampling helps us separate 1-halo and 2-halo terms when modeling the total galaxy power spectrum in redshift space:
\begin{equation}
	P_{gg}(\bold{k}) = (1-f_s)^2 P_{cc}(\bold{k}) + 2f_s(1-f_s)P_{cs}(\bold{k}) + f_s^2P_{ss}(\bold{k}),
\end{equation}
where $P^{cc}, P^{cs}$, and $P^{ss}$ are the auto-power spectrum of centrals, the central-satellite cross-power spectrum, and the auto-power spectrum of satellites, respectively, and $f_s$ is the satellite fraction.
We also account for non-linear distortions caused by the large virial motions of satellite galaxies within their halos - known as the Finger-of-God effect. We model this effect with a Lorentzian damping factor $G(k\mu;\sigma_v)=(1+k^2\mu^2\sigma_v^2/2)^{-2}$ applied to the redshift-space power spectrum of each sub-sample, where $\sigma_v$ is the velocity dispersion of the sample. In this model, we assume a single velocity dispersion parameter for both type A and type B centrals, $\sigma_c$, and parameters for type A satellites, $\sigma_{s_A}$, and for type B satellites, $\sigma_{s_B}$. We take $\sigma_c$ and $\sigma_{s_A}$ as free parameters and determine $\sigma_{s_B}$ from the relation between the linear bias and velocity dispersion, using the relation for the halo mass and bias and the virial theorem scaling between velocity dispersion and mass.

The resulting galaxy power spectrum model depends on the following 11 physically-motivated parameters:
$$[f(z_{\mathrm{eff}}), \sigma_8(z_{\mathrm{eff}}), b_{1, c_A}, b_{1, s_A}, b_{1, s_B}, f_s, f_{s_B}, \langle N_{>1,s}\rangle, \sigma_c, \sigma_{s_A}, f^{1h}_{s_B s_B}].$$
This includes two cosmological parameters, the growth rate $f$ and the amplitude of matter fluctuations $\sigma_8$ evaluated at the effective redshift of the sample $z_{\mathrm{eff}}$, and linear bias parameters of type A centrals and type A and B satellites ($b_{1, c_A}$, $b_{1, s_A}$, $b_{1, s_B}$). We also consider the fraction of all satellites $f_s$, the fraction of type B satellites $f_{s_B}$, and the mean number of satellite galaxies in halos with more than one satellite $\langle N_{>1,s}\rangle$. The velocity dispersion parameters for some sub-samples ($\sigma_c$ and $\sigma_{s_A}$) are accounted for, and we also vary the normalization nuisance parameter for the 1-halo amplitude $f^{1h}_{s_B s_B}$. Additionally, there exists the Alcock-Paczynski (AP) effect, geometric distortions of the galaxy statistics due to the mismatch between the true cosmology and the fiducial cosmology which we assumed when converting redshifts and angular positions of the observed galaxies into the three-dimensional physical positions.
The AP parameter  is
very insensitive to cosmological
parameters
\cite{beutler20126df, BeutlerEtAl17}. As a result
we fix the scaling factors $\alpha_{\parallel}$ and $\alpha_{\perp}$ of the AP effect to their fiducial values
of Planck cosmology. 
We have verified that varying 
$F_{AP}$ within its error from Planck+BAO has 
negligible effect on the 
final constraints. 
The primary
goal of this
paper is to determine the growth rate
$f \sigma_8$, and we will not attempt
to separate $\sigma_8$ from $f$.

Since we only wish to determine
one parameter $f \sigma_8$, 
the question remains whether 
fixing other parameters that 
modify the shape of the power
spectrum has an impact on its
value or its error. We have tested this by performing the 
full analysis of a set of 
30 power spectra drawn 
at random from Planck+BAO 
chains. We find a 
scatter in $f\sigma_8$ best 
fit value at the 
level of roughly 30\% of the 
statistical error. When 
combining errors in quadrature 
this leads to an additional 
5\% increase in statistical error, which 
we include in the final 
results. 

\subsection{Measurement: power spectrum estimator}\label{3.2}

We measure the galaxy clustering signal via the multipole moments of the power spectrum $P_l(k)$, using \texttt{nbodykit}, the python software package for large-scale structure data analysis. In \texttt{nbodykit}, the FFT-based algorithm for the anisotropic power spectrum estimator, presented in \cite{Hand:2017irw}, is implemented. This provides fast evaluation of the estimator in \cite{Yamamoto:2005dz} by expanding the Legendre polynomials into spherical harmonics rather than using a Cartesian decomposition and thereby requiring only $2l+1$ FFTs to obtain a multipole of order $l$.


We estimate the power spectrum multipoles as:
\begin{equation}
	P_l(k) = \frac{2l+1}{A}\int \frac{d\Omega_k}{4\pi}F_0(\bold{k})F_l(\bold{-k}),
\end{equation}
where $\Omega_k$ is the solid angle in Fourier space, and $\mathcal{L}_l$ is the Legendre polynomial. $A$ is the normalization defined as $A \equiv \int d\bold{r} [n^{\prime}_{\mathrm{gal}}(\bold{r})w_{\mathrm{fkp}}(\bold{r})]^2$, where $n^{\prime}_{\mathrm{gal}}$ is the weighted galaxy number density field, and $w_{\mathrm{fkp}}$ is the FKP weight.

The weighted galaxy density field $F(\bold{r})$ is given by
\begin{equation}
	F(\bold{r}) = \frac{w_{\mathrm{fkp}}(\bold{r}) }{A^{1/2}}[n^{\prime}_{\mathrm{gal}}(\bold{r}) - \alpha^{\prime} n^{\prime}_{\mathrm{ran}}(\bold{r})],
\end{equation}
where $n^{\prime}_{\mathrm{gal}}$ and $n^{\prime}_{\mathrm{ran}}$ are the number density field for the galaxy and randoms catalogues respectively, with the factor $\alpha^{\prime}$ normalizing $n^{\prime}_{\mathrm{ran}}$ to $n^{\prime}_{\mathrm{gal}}$, and
\begin{align}
\begin{split}
F_l(\bold{k}) &= \int d\bold{r}F(\bold{r})e^{i\bold{k}\cdot\bold{r}}\mathcal{L}_l(\bold{\hat{k}}\cdot\bold{\hat{r}}) \\
&= \frac{4\pi}{2l+1}\sum_{m=-l}^{l}Y_{lm}(\bold{\hat{k}})\int d\bold{r}F(\bold{r})Y_{lm}^*(\bold{\hat{r}})e^{i\bold{k}\cdot\bold{r}}.
\end{split}
\end{align}
We compute each summation over $m$ using a FFT, hence a total of $2l+1$ FFTs.

\begin{figure}[t]
\centering
\includegraphics[width=.48\textwidth]{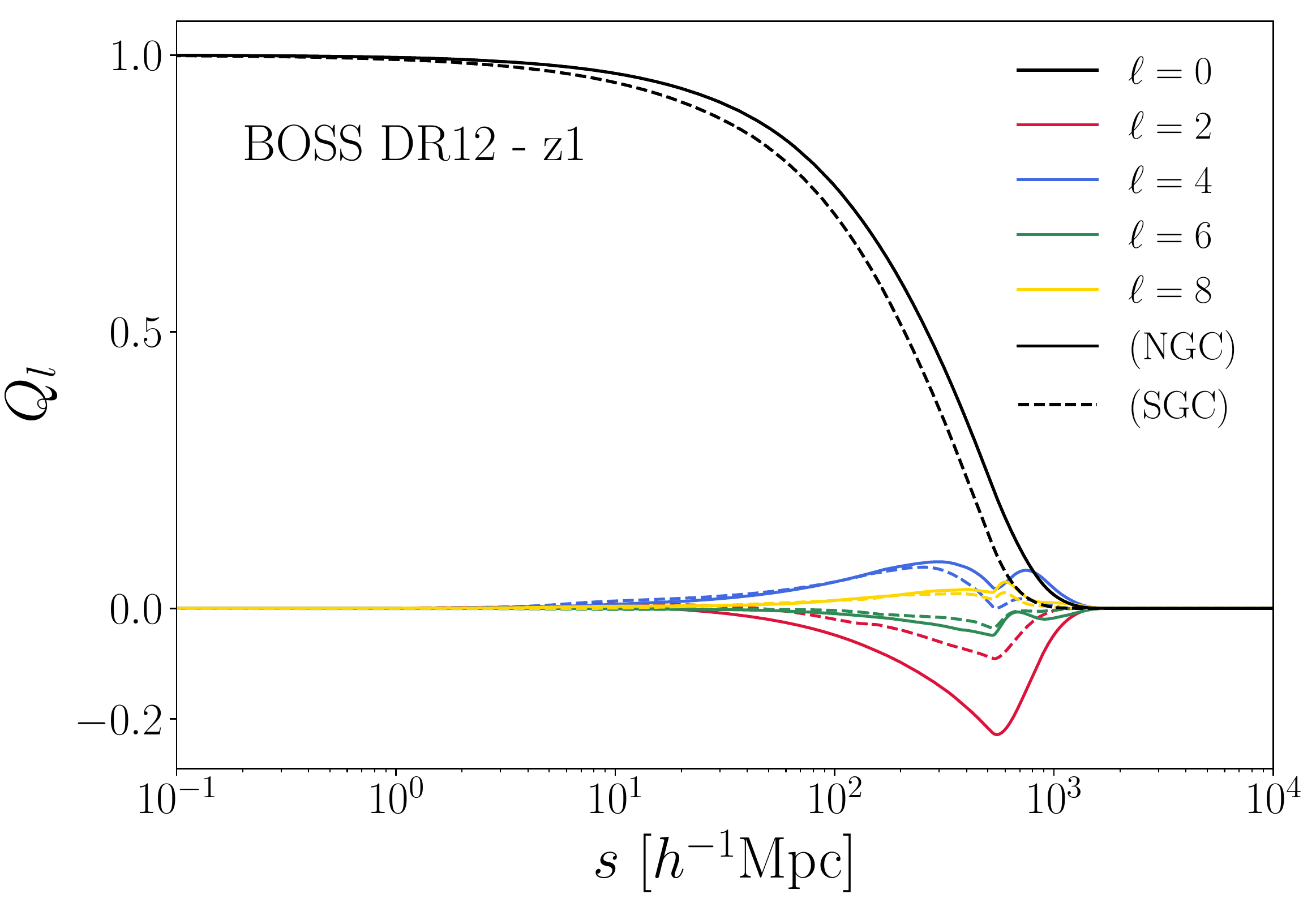}
\hfill
\includegraphics[width=.48\textwidth]{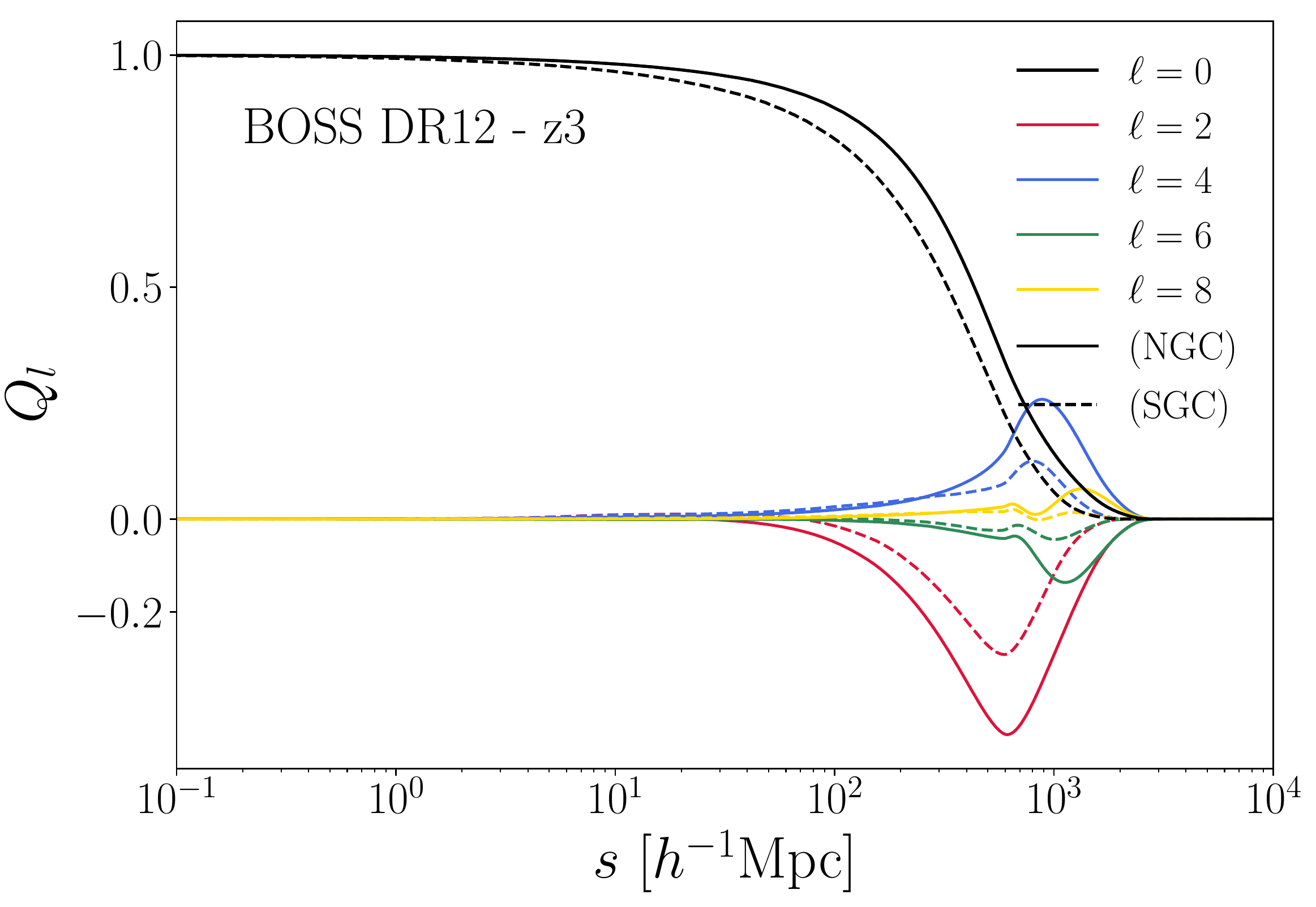}
\caption{\label{fig:WF1} The window function multipoles in configuration space for the BOSS DR12 z1 (\textit{Left}) and z3 (\textit{Right}) samples. We include up to $Q_8(s)$ because the contribution of $l=10$ or higher is negligible for the window convolution.}
\end{figure}

However, \cite{Beutler:2021eqq} replaces the traditional definition of the normalization term $A$ with the value enforcing the following condition on the window function multipole of order $l=0$, $Q_0(s \rightarrow 0)=1$, to ensure that the power spectrum and window function are normalized in a consistent way. Table 1 in \cite{Beutler:2021eqq} shows that such correction of the normalization term results in increasing the BOSS DR12 galaxy power spectrum multipole amplitudes by roughly 10\%. Following \cite{Beutler:2021eqq}, we corrected all galaxy power spectrum measurements presented in this work.

\subsection{Survey window function}

We account for the window function effects by convolving the theoretical model in section~\ref{3.1} with the survey window function, which corresponds to the Fourier transform of the survey volume. We denote the resulting quantity as the ``convolved'' power spectrum.

\begin{figure}[t]
\centering
\includegraphics[scale=0.3]{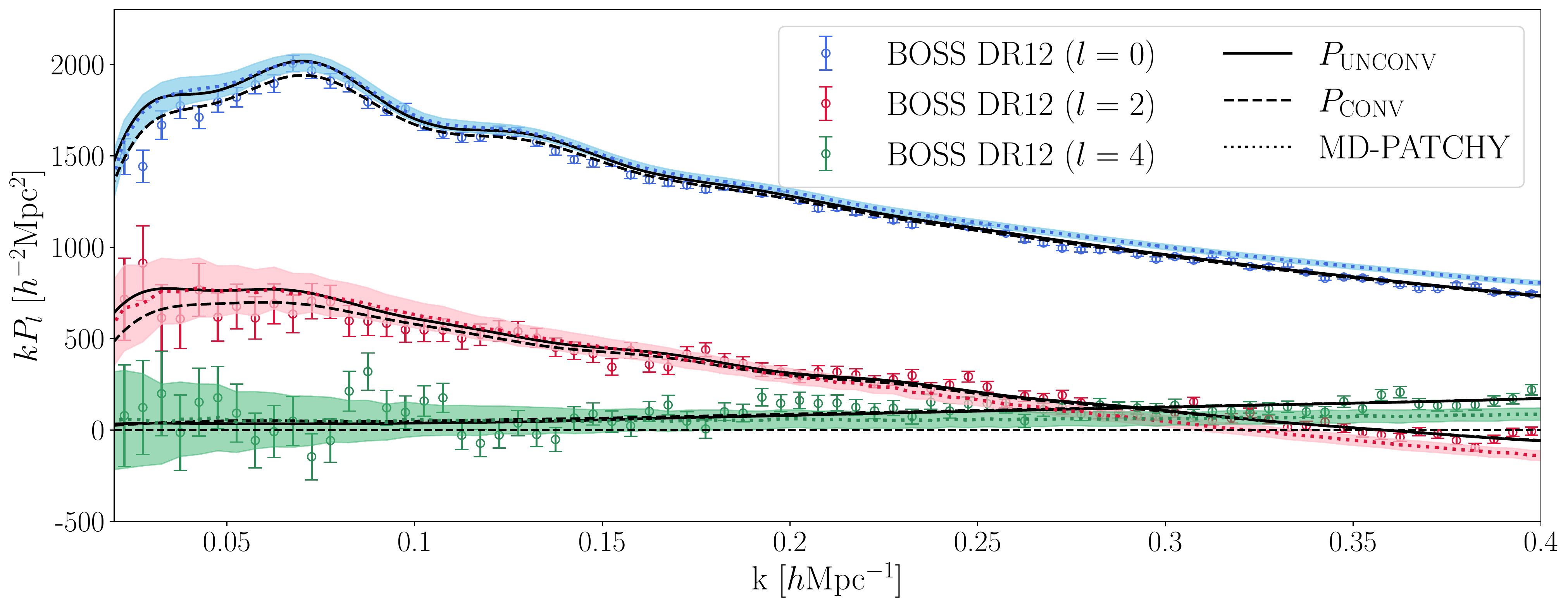}
\caption{The window function effects on the power spectrum multipoles (filled circles with error bars) for the BOSS DR12 z3 NGC sample. The solid and dashed curves correspond to the unconvolved and convolved multipoles, respectively. We also show the mean of 1000 MD-PATCHY V6C mock catalogues (colored dotted curves), and shaded areas indicate 1$\sigma$ deviations of 1000 mocks. Dotted curves match well with the data points, suggesting that V6C catalogues reproduce the clustering of the observed data (which makes them suitable for the covariance matrix estimation). For our main analysis, we choose the minimum wavenumber of $k_{\mathrm{min}}=0.02 h$Mpc$^{-1}$ to minimize any large-scale effects
of the window function.}
\label{fig:WF2}
\end{figure}

We follow the method presented in \cite{WilsonPeacockEtAl15} to compute the convolved power spectrum multipoles. First, we use the pair counting algorithm in \texttt{nbodykit} \citep{nbodykit}, which employs the \texttt{Corrfunc} package \citep{corrfunc}, to obtain the pair counts  of the random catalogue. In figure~\ref{fig:WF1}, we present the resulting window function multipoles, $Q_l(s) \propto \int_{-1}^{1}d\mu RR(s,\mu)\mathcal{L}_l(\mu) \approx \sum_i RR(s_i, \mu_i)\mathcal{L}_l(\mu_i)$, of the BOSS DR12 galaxy samples in configuration space, for a set of separations $s$. The z1 and z3 window function multipoles vanish on scales larger than $\approx$ 2000 and 3000 $h^{-1}$Mpc, respectively, and these correspond to the largest scales in the volume of the BOSS galaxy samples. In this work, we ignore $l=10$ or higher, as its contribution to the convolution is negligible.

Next, we convolve the correlation function multipoles $\xi_l(s)$ obtained from the theoretical model with the survey window function to get the convolved multipoles $\hat{\xi}(s)$:
\begin{align}
\begin{split}
\hat{\xi}_0 = & \xi_0 Q_0 + \frac{1}{5}\xi_2 Q_2 + \frac{1}{9}\xi_4 Q_4 + ... \\
\hat{\xi}_2 = & \xi_0 Q_2 + \xi_2 \Big[ Q_0 + \frac{2}{7}Q_2 + \frac{2}{7}Q_4 \Big] \\
& + \xi_4 \Big[ \frac{2}{7}Q_2 + \frac{100}{693}Q_4  + \frac{25}{143}Q_6 \Big] + ... \\
\hat{\xi}_4 = & \xi_0 Q_4 + \xi_2 \Big[ \frac{18}{35}Q_2 + \frac{20}{77}Q_4 + \frac{45}{143}Q_6 \Big] \\
& + \xi_4 \Big[ Q_0 + \frac{20}{77}Q_2 + \frac{162}{1001}Q_4  + \frac{20}{143}Q_6  + \frac{490}{2431}Q_8 \Big] + ...
\end{split}
\end{align}
Figure~\ref{fig:WF2} demonstrates that the effects of the window function is mostly on large scales. In this work, we choose the minimum wavenumber as $k_{\mathrm{min}}=0.02 h$Mpc$^{-1}$ to minimize any large-scale effects of the window function.

\section{Analysis methods}\label{4}

\subsection{Covariance matrices}\label{sec4}

Fitting the theoretical model to the measured data requires a covariance matrix estimate, and in this work we not only consider the covariance matrix from mock catalogues (section~\ref{4.1}) but also the hybrid covariance matrix which combines the analytic disconnected component (\ref{4.2}) and smoothed connected component (\ref{4.3}).

\subsubsection{Mock covariance matrix}\label{4.1}

We compute the covariance matrix from 1000 realizations of MD-PATCHY mock catalogues (section~\ref{2.2}):
\begin{align}
\mathrm{Cov}\Big[P_{l}(k_i),P_{l^{\prime}}(k_j)\Big] =& \frac{1}{N-1} \sum_{\alpha=1}^N \Big[ P_{l,\alpha}(k_i) - \overline{P}_l(k_i) \Big] \cdot \Big[ P_{l^{\prime},\alpha}(k_j) - \overline{P}_{l^{\prime}}(k_j) \Big],
\end{align}
where N is the number of mocks, and $\overline{P}_l(k) = \frac{1}{N}\sum_{\alpha=1}^N P_{l,\alpha}(k)$ is the mean power spectrum. Hence, we obtain the covariances between multipoles (for $l=0,2,4$) along with their uncertainties. For all multipoles, the fitting range is $0.02 < k < 0.4\  h$Mpc$^{-1}$ and $\Delta k = 0.005\ h$Mpc$^{-1}$  (corresponding to $N_{\mathrm{bin}} = 76$ for each $l$). We also apply the Hartlap correction \citep{HartlapSimonEtAl07} to get an unbiased estimate of the true inverse covariance matrix.

\subsubsection{Analytic disconnected covariance matrix}\label{4.2}
Following \cite{Li2019}, we compute the analytic ``disconnected'' covariance matrix (more conventionally, ``Gaussian'' covariance matrix) which takes into account the window effect. This analytic method is free of sampling noise and therefore avoids numerical issues of the mock covariance matrix.

Assuming the flat sky approximation, we write the ensemble average of the estimated power spectrum as
\begin{equation}
    \ensavg{\hat{P}(\vk)} = \frac1{\W_0} \int_\vq P(\vk - \vq) |W(\vq)|^2 \simeq \frac{P(\vk)}{\W_0} \int_\vq |W(\vq)|^2\ (\mathrm{for}\ k \gg q) = P(\vk),
    \label{Phat_expect}
\end{equation}
where $W(\vx) \equiv \bar{n}_{\mathrm{gal}}(\vx) w(\vx)$ denotes the windows on the fields, with $\bar{n}_{\mathrm{gal}}(\vx) = \ensavg{n_{\mathrm{gal}}(\vx)}$ and the weight $w(\vx)$ which minimizes systematic effects. This shows that we get the ensemble average of the estimator $\hat{P}$ by convolving the true power spectrum $P$ with a window, and $\hat{P}$ is an unbiased estimate of $P$ for scales much smaller than the window. We then split the covariance $\cov\bigl[\hat{P}(\vk), \hat{P}(\vk')\bigr]
    = \ensavg{\hat{P}(\vk) \hat{P}(\vk')}
    - \ensavg{\hat{P}(\vk)} \ensavg{\hat{P}(\vk')}$ into the disconnected and connected pieces:
\begin{align}
\cov\bigl[\hat{P}(\vk), \hat{P}(\vk')\bigr]
    &= \cov^\disc\bigl[\hat{P}(\vk), \hat{P}(\vk')\bigr]
    + \cov^\conn\bigl[\hat{P}(\vk), \hat{P}(\vk')\bigr].
\end{align}
The disconnected covariance component $\cov^\disc$ involves quadratic combinations of the following window factors which modulate Gaussian and Poisson parts: $\W(\vq) = \int_\vx \W(\vx) e^{-i\vq\cdot\vx} \equiv \int_\vx W(\vx)^2 e^{-i\vq\cdot\vx}$ and $\S(\vq) = \int_\vx \S(\vx) e^{-i\vq\cdot\vx} \equiv (1 + \alpha) \int_\vx \nbar(\vx) w(\vx)^2 e^{-i\vq\cdot\vx}$. We further define the factor $\QW, \QS$ and $\QX$ as the auto- and cross-correlations of $\W$ and $\S$,
\begin{align}
    \QW(\vq) \equiv \W(\vq) \W(\vq)^* &= \int_\vs \QW(\vs) e^{-i\vq\cdot\vs},
    \nonumber\\
    \QS(\vq) \equiv \S(\vq) \S(\vq)^* &= \int_\vs \QS(\vs) e^{-i\vq\cdot\vs},
    \nonumber\\
    \QX(\vq) \equiv \W(\vq) \S(\vq)^* &= \int_\vs \QX(\vs) e^{-i\vq\cdot\vs},
    \label{Q}
\end{align}
and write the disconnected covariance as
\begin{multline}
    \cov^\disc\bigl[\hat{P}(\vk), \hat{P}(\vk')\bigr]
    \approx \frac1{\W_0^2} \Bigl\{ P(\vk) P(\vk') \QW(\vk-\vk')
    \\
    + \bigl[P(\vk) + P(\vk')\bigr] \Re\bigl[\QX(\vk-\vk')\bigr]
    + \QS(\vk-\vk') \Bigr\}
    + (\vk' \leftrightarrow -\vk').
    \label{disc_P}
\end{multline}
We refer the reader to \cite{Li2019} for a complete and detailed derivation.

\begin{figure}[tbp]
\centering
    \includegraphics[width=0.45\textwidth]{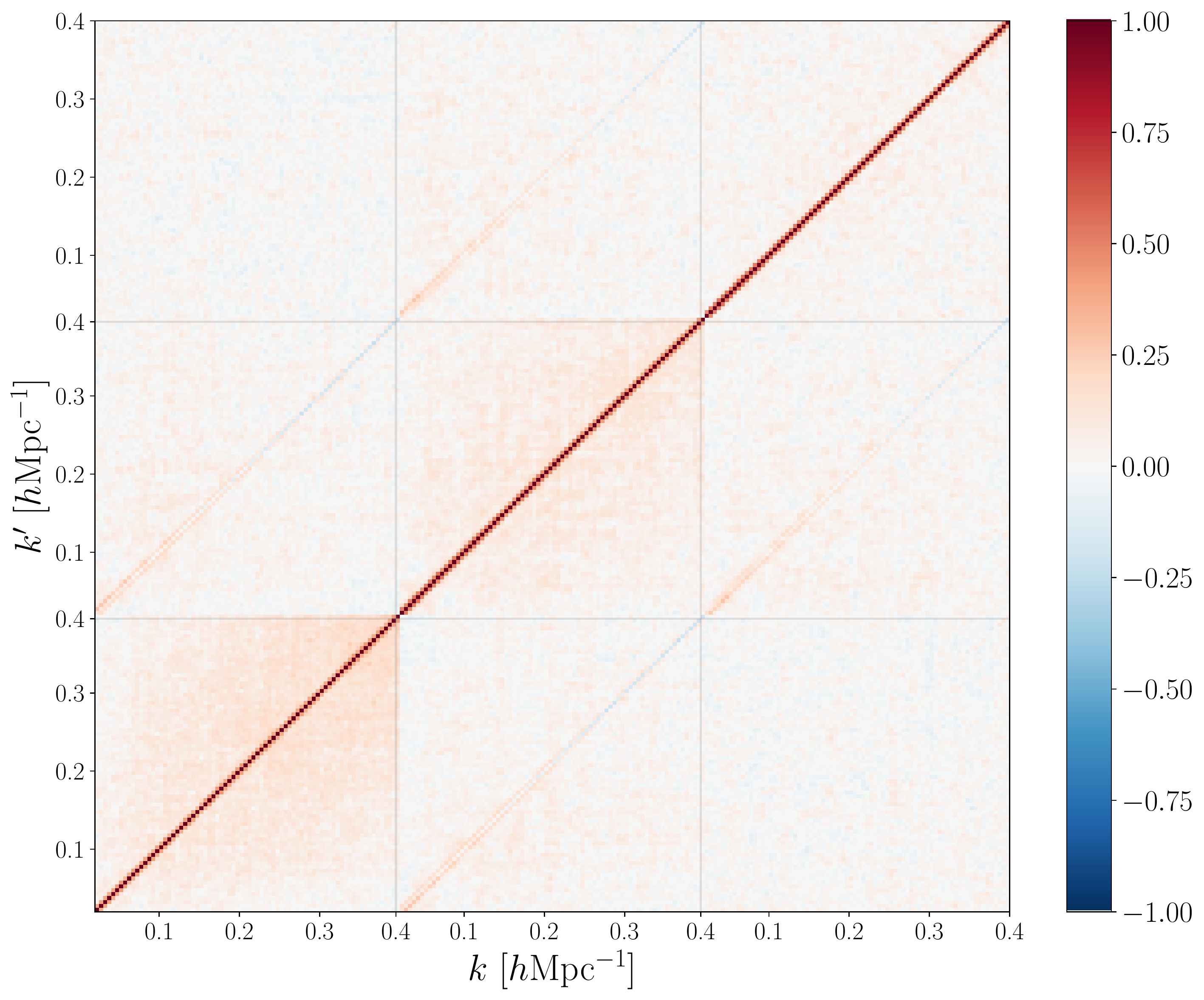}
    \includegraphics[width=0.45\textwidth]{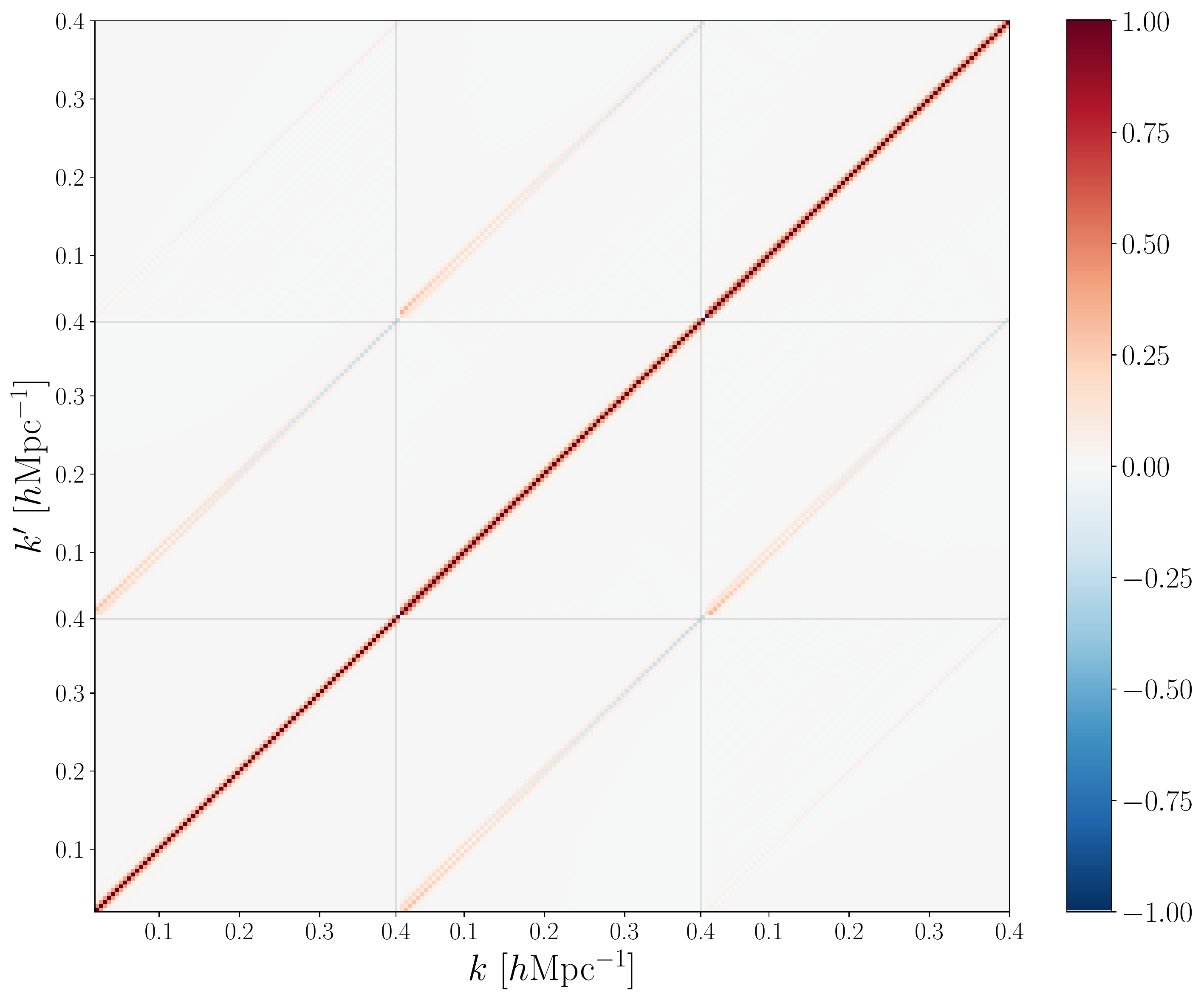}
    \caption{3 $\times$ 3 blocks of correlation matrices of the power spectrum multipoles, which visualize the auto- and cross-correlations of $P_0, P_2$ and $P_4$. \textit{Left}: Mock correlation matrix from 1000 MD-PATCHY z3 NGC mock simulations (section~\ref{4.1}). \textit{Right}: Analytic disconnected correlation matrix for the z3 NGC sample (section~\ref{4.2}).}
    \label{fig:cov1}
\end{figure}

Figure~\ref{fig:cov1} presents the 3 $\times$ 3 blocks of correlation matrices from mock catalogues (left panel) and from the analytic method described in this section (right panel). Correlation coefficients are defined as
\begin{align}
    \mathrm{Corr}(O,O^{'}) = \frac{\cov(O,O^{'})}{\sqrt{\cov(O,O)\cov(O^{'},O^{'})}},
\end{align}
where $O \in \{ P_0(k), P_2(k), P_4(k) \}$ and $O^{'} \in \{ P_0(k^{'}), P_2(k^{'}), P_4(k^{'}) \}$. \cite{Li2019} demonstrates that the analytic Gaussian covariance matrix is in excellent agreement with the mock covariance matrix, and this method is free of the sampling noise with much smaller computational cost, compared to the mocks.

\begin{figure}[t]
\centering
    \includegraphics[scale=0.3]{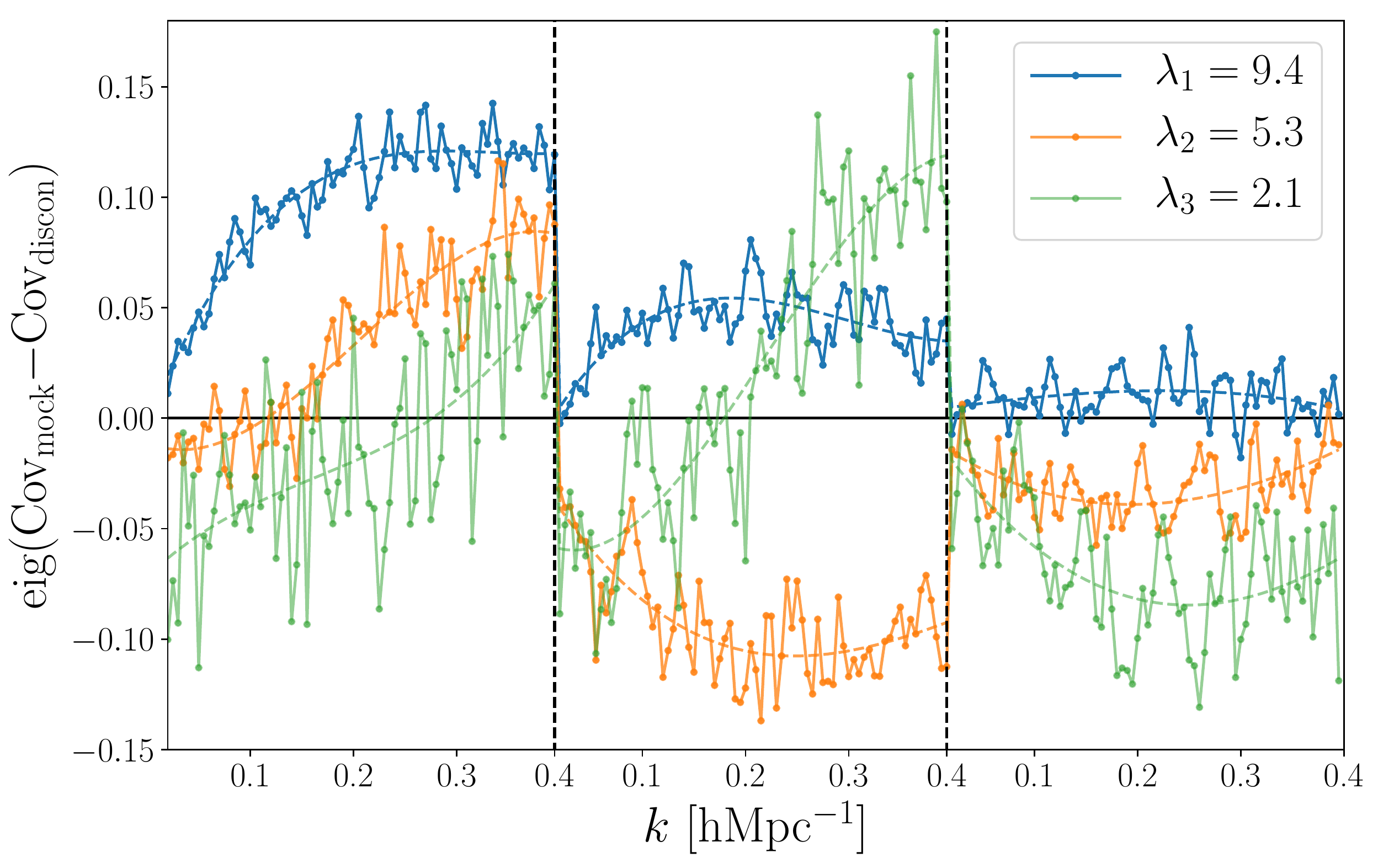}
    \includegraphics[scale=0.3]{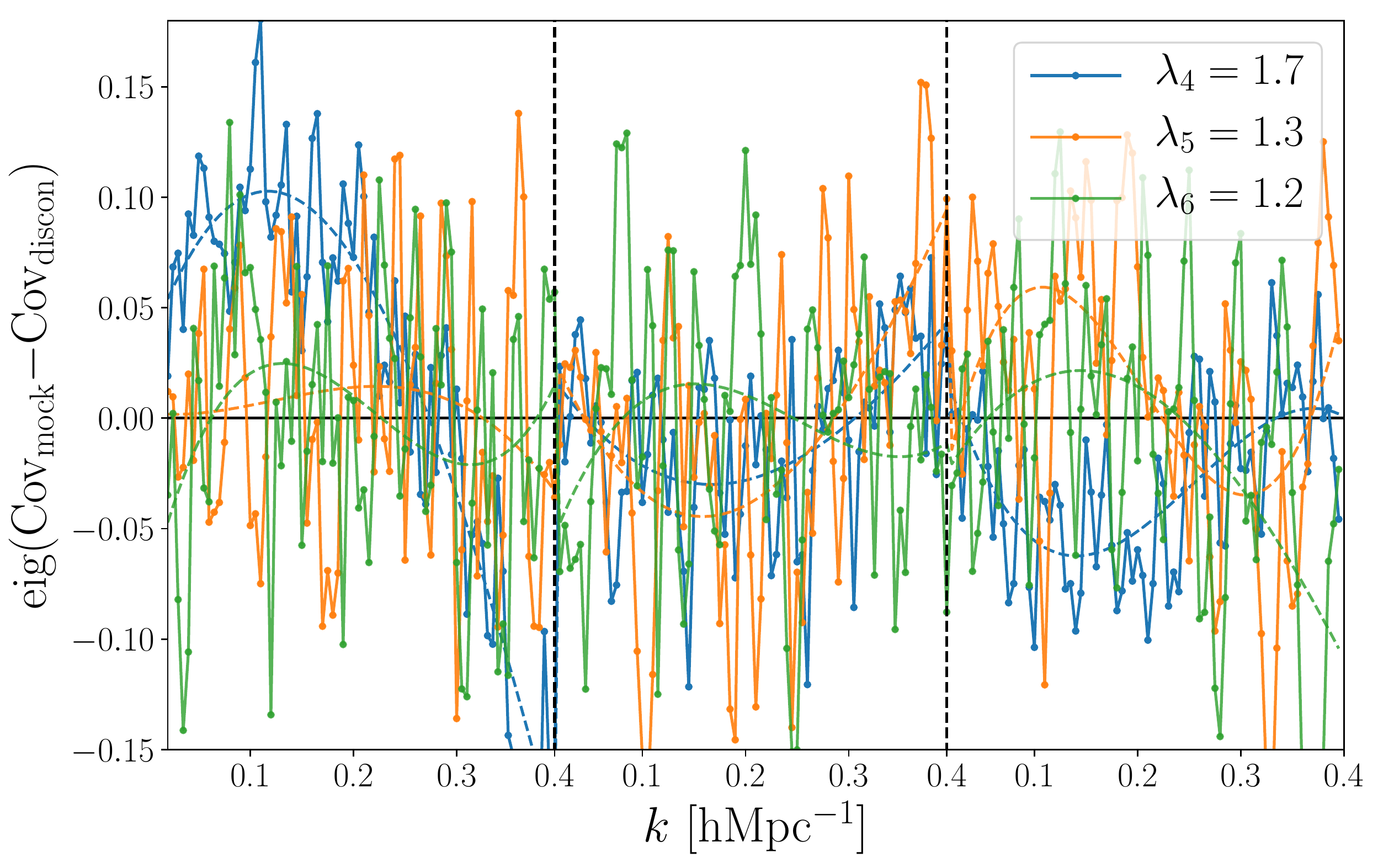}
    \caption{The first six principal components of the $(3 \times 3)$ blocks of $P_l$ from 1000 MD-PATCHY z3 NGC mock simulations, where $l = 0,2,4$. We only include up to four principal components for a low-rank approximation, as the components beyond the fourth are noisy and do not contain much broadband correlations. $\lambda_i$ denotes the eigenvalue of the $i$-th eigenvector.}
    \label{fig:cov2}
\end{figure}

\begin{figure}[t]
\centering
\includegraphics[width=0.45\textwidth]{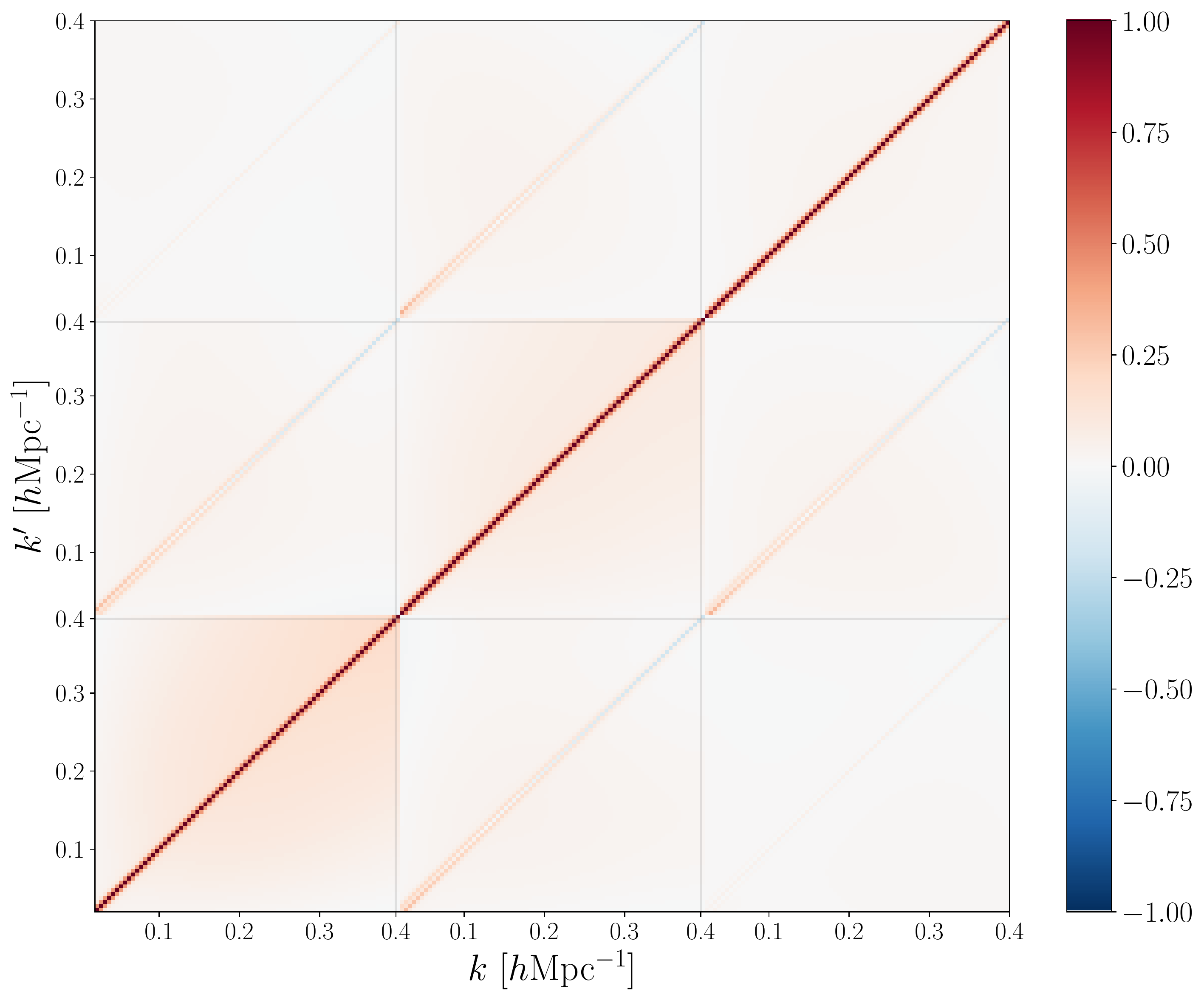}
\includegraphics[width=0.45\textwidth]{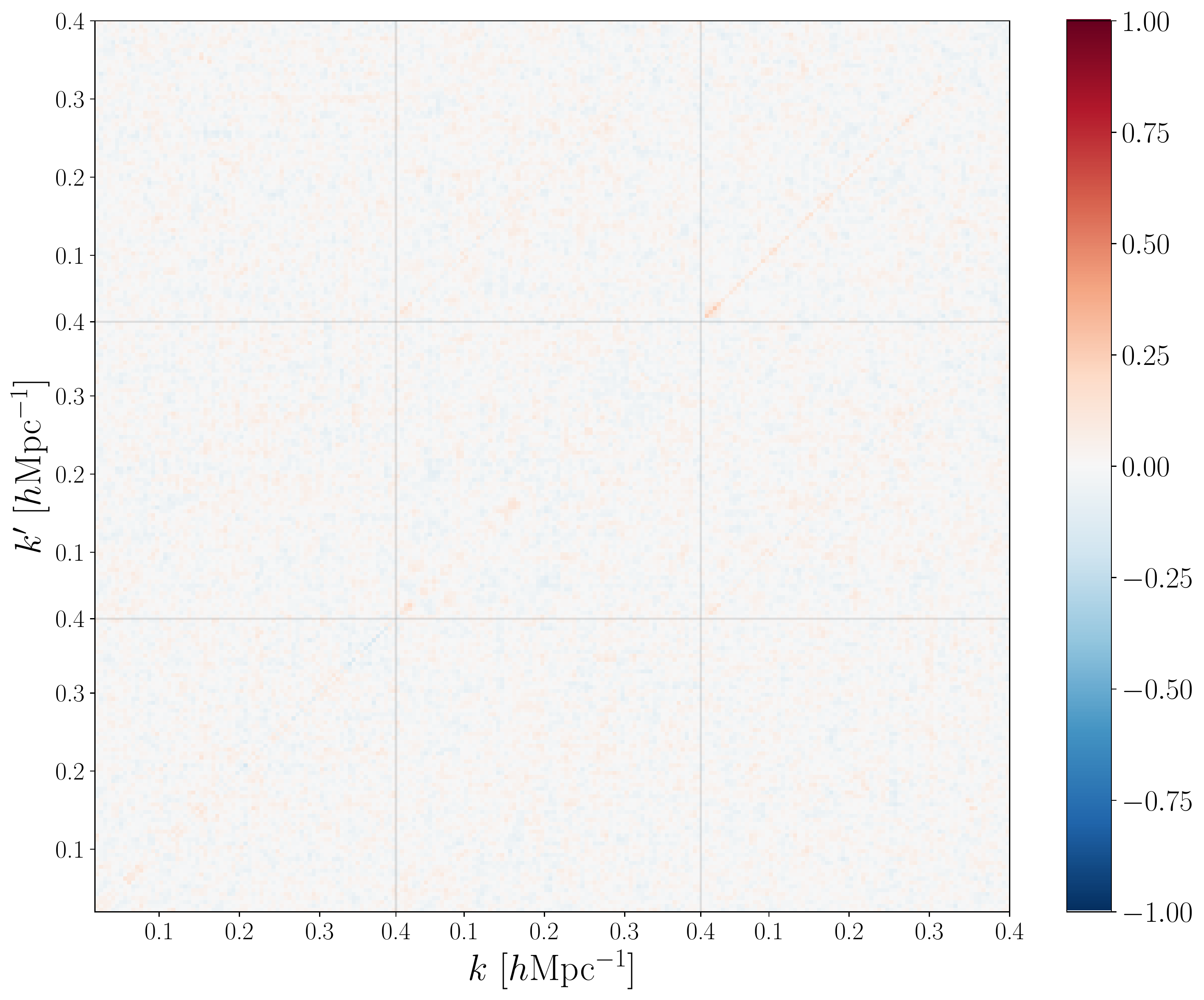}
\caption{\textit{Left}: we combine the analytic Gaussian covariance with the smoothed connected pieces obtained from low-rank components with principal component analysis. \textit{Right}: The difference between the mock and the hybrid covariance which includes the analytic disconnected and smoothed connected covariances, normalized by the diagonal of the latter. Because the connected part is smooth and has a low-rank approximation, we find that only the first four principal components are needed to obtain a smooth estimate of the connected part, and the difference between the mock and its smooth estimate is clean.}
\label{fig:cov_diff}
\end{figure}

\subsubsection{Modeling the connected pieces using PCA}\label{4.3}
The remaining connected pieces of the covariance matrix, which include Poisson, non-Gaussian (trispectrum) and Gaussian (power spectrum) components, is more difficult to model analytically than the disconnected piece. However, an eigenmode decomposition of the connected components shows that it is a low-rank matrix \citep{HarnoisPen12, MohammedSeljakVlah17, Wadekar:2020hax}. In this work, we first obtain the connected parts from the MD-PATCHY mock simulations by subtracting the disconnected parts from the mock covariance matrix and show that this empirical connected piece is indeed a low-rank component with principal component analysis (PCA).

In Figure~\ref{fig:cov2}, we perform PCA on the $(3 \times 3)$ blocks of $P_l$, where $l = 0,2,4$, and show its first six principal components. The principal components beyond the fourth component are noisy and do not carry much broadband correlations, and therefore we only include up to four principal components; therefore, the connected components can be well approximated by a low-rank eigen-decomposition. The resulting smoothed connected covariance, combined with the analytic Gaussian covariance, is shown on the left panel of Figure~\ref{fig:cov_diff}, and the right panel shows that it achieves good agreement with the mocks.

\subsection{Parameter estimation techniques}\label{5}

In this work, we use the following methods to obtain the parameter posterior distribution: 1) Maximum a posteriori (MAP) estimation and Laplace approximation, using the hessian of the log posterior to obtain the inverse covariance matrix of the model parameters and 2) MCMC sampling of the likelihood, assuming the hybrid covariance matrix with the smoothed connected parts. In section~\ref{6.0}, we find the best-fitting model parameters for each of the 1000 mock catalogues from MAP estimation and present the 1D histograms and 2D correlations of the cosmological parameters of our interest. Section~\ref{7} summarizes the main results of this work: BOSS DR12 RSD measurements of the growth of structure, and we obtain the parameter posteriors using the Python module \texttt{emcee} \cite{ForemanHoggEtAl13}. \cite{Seljak:2019rta} presents an optimization-based posterior inference method called EL$_2$O and shows that the posterior distribution from EL$_2$O agrees with the MCMC results. Particularly, section 4.4 in \cite{Seljak:2019rta} discusses how EL$_2$O can be effective in galaxy clustering analyses.
We refer the reader to \cite{Seljak:2019rta} for more detailed analysis about the comparison between EL$_2$O and MCMC methods.

\section{Model Performance}\label{6}


\begin{figure}[tbp]
\centering
    \includegraphics[scale=0.243]{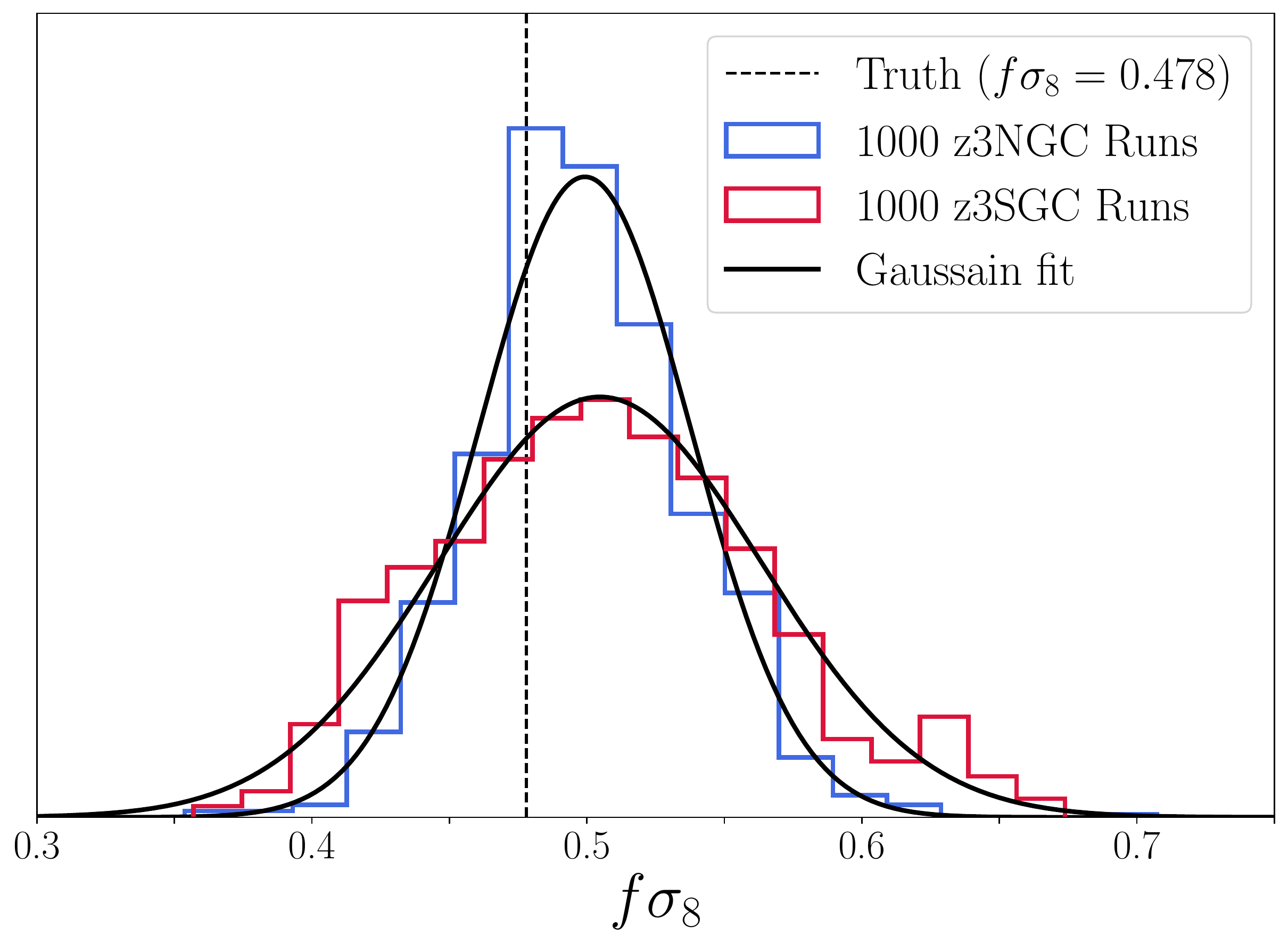}
    \includegraphics[scale=0.273]{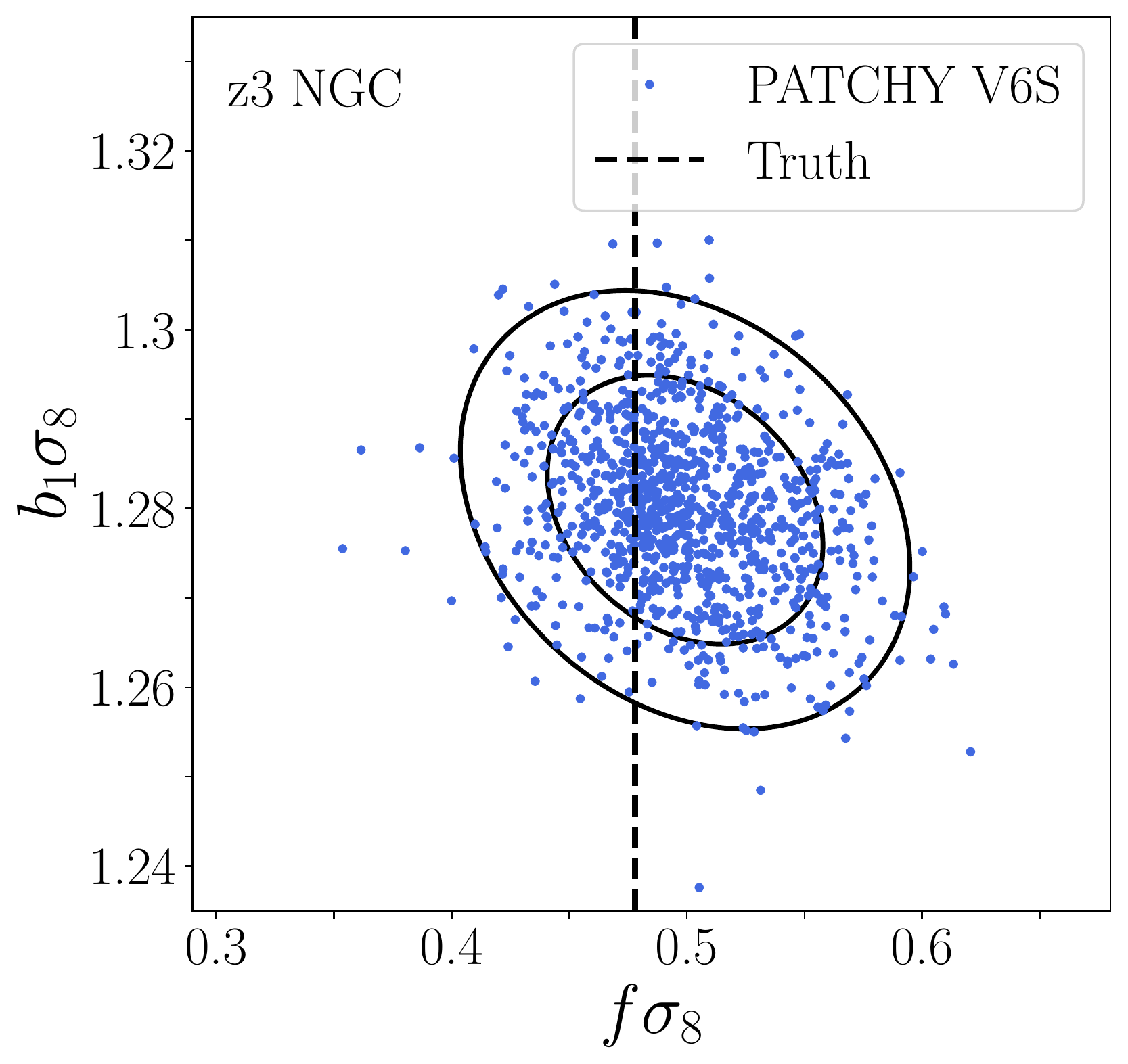}
    \includegraphics[scale=0.273]{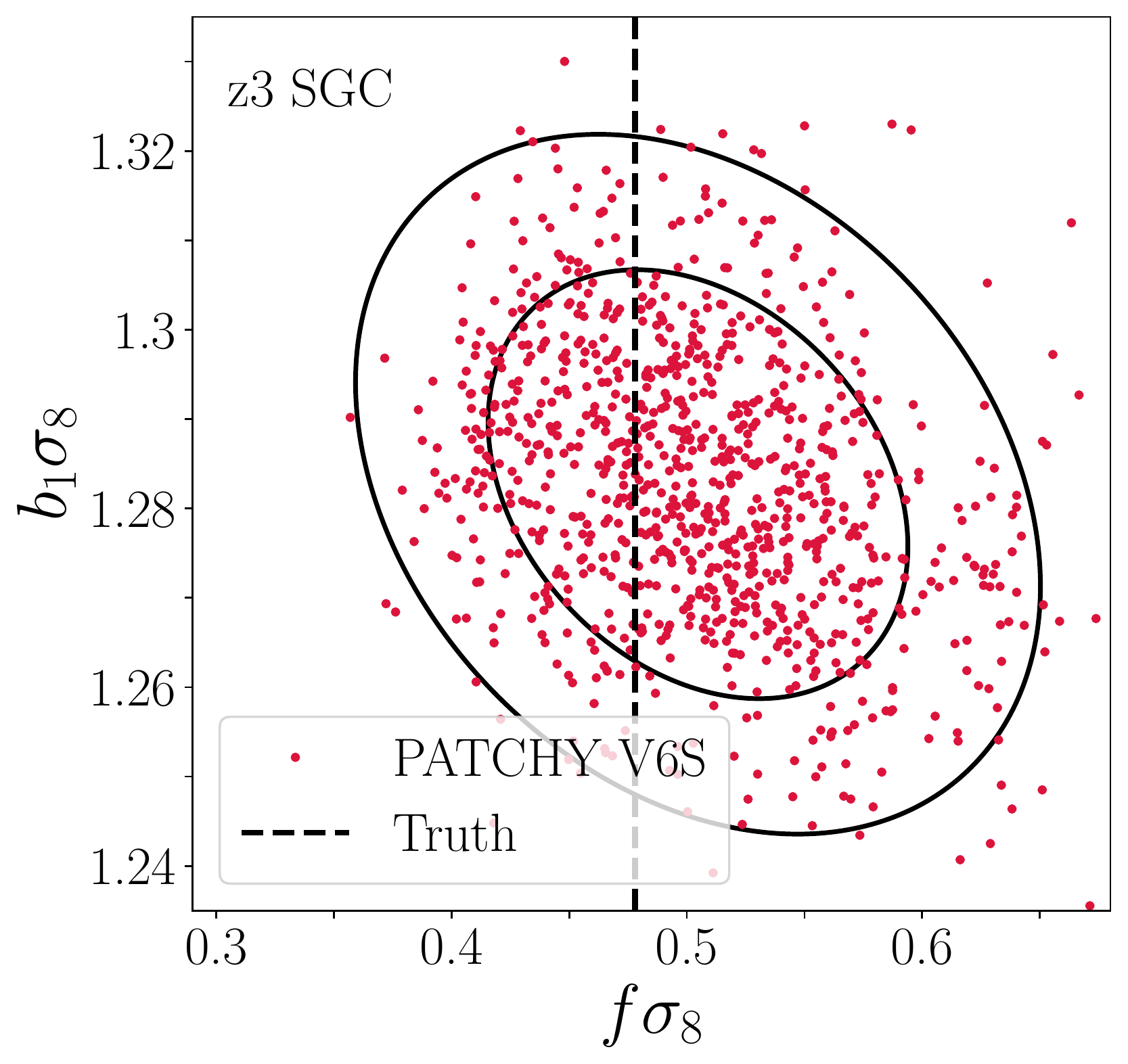}
    \caption{ \textit{Left}: Maximum a posteriori (MAP) results for 1000 MD-PATCHY z3 NGC (blue) and SGC (red) mock catalogues. We fit the monopole, quadrupole, and hexadecapole over the wavenumber range $0.02 < k < 0.2\ h$Mpc$^{-1}$. 1D distribution of our $f\sigma_8$ fit results gives $f\sigma_8=0.499 \pm 0.038$ and $f\sigma_8=0.502 \pm 0.058$, for z3 NGC and SGC respectively. The true cosmology indicates $f\sigma_8=0.478$,
    which is 0.4-0.5$\sigma$ away from the mean of our recovered values. \textit{Middle, Right}: 2D correlations of $f\sigma_8$ and $b_1\sigma_8$ for 1000 MD-PATCHY z3 NGC (blue) and SGC (red) mock catalogues. Vertical dashed line indicates the true cosmology, and solid contours show 1$\sigma$ and 2$\sigma$ confidence regions.
    }
    \label{fig:PATCHYfit}
\end{figure}

\subsection{Tests on the MultiDark-PATCHY mocks}\label{6.0}

In \cite{Hand:2017ilm}, the accuracy and precision of the power spectrum model in section~\ref{3.1} are extensively assessed by performing independent tests using several sets of mocks based on high-fidelity, periodic $N$-body simulations and realistic BOSS DR12 CMASS mocks. The fits typically have a low bias 
up to $k=0.4$h/Mpc, although for fits with 
free AP parameter the bias is in some cases 
significant. We argue in the introduction that 
this could be caused by a small model 
misspecification, which can be amplified whenever
one must break the degeneracy between two 
strongly correlated parameters. For this 
reason we fix AP parameter in this paper. 
To further confirm that this model is unbiased and accurate enough to provide cosmological parameter constraints of the BOSS DR12 sample, we fit our model to 1000 MD-PATCHY mock catalogues and verify that we can retrieve the true cosmology, provided by the BigMultiDark simulation.

Applying the analysis pipeline in section~\ref{5} to the mock catalogues, we obtain the best-fitting parameters for each of the 1000 catalogues by first measuring the power spectrum multipoles ($l=0,2,4$) for each of the catalogues and obtaining the MAP estimate using the L-BFGS algorithm. Figure~\ref{fig:PATCHYfit} presents the MAP results for 1000 MD-PATCHY z3 NGC (blue) and SGC (red) mock catalogues, and the black dotted vertical line indicates the expected parameter value from the true cosmology of the MD-PATCHY simulations. If we include up to $k_{\mathrm{max}}=0.2\ h$Mpc$^{-1}$, we find that $f\sigma_8=0.499 \pm 0.038$ for z3 NGC and $f\sigma_8=0.502 \pm 0.058$ for z3 SGC, so we recover the true cosmology ($f\sigma_8=0.478$) within 1$\sigma$ with only modest mean biases of $\Delta f\sigma_8$ of 0.5$\sigma$.

\subsection{Choice of $k_{\mathrm{max}}$}\label{6.1}

We can further investigate whether the
0.4-0.5 $\sigma$ bias we observe
in Figure \ref{fig:PATCHYfit}
is caused by the
priors or by model misspecification, by
comparing it to the analysis where we
treat all 1000 mocks as a single dataset.
In this case the MAP will be dominated
by the likelihood and priors can be
ignored. We find $f\sigma_8=0.478$ for z1 NGC and $f\sigma_8=0.490$ for z3 NGC for
$k_{\mathrm{max}}=0.2\ h$Mpc$^{-1}$,
compared to the truth ($f\sigma_8=0.484$ for z1 and 0.478 for z3), both with very small error of order 0.001. In both cases the mean
has moved closer to the true value,
suggesting the prior is driving the
MAP away from the true value, even
if there is also some small
model misspecification for z3 NGC. 
If we
repeat the analysis for
$k_{\mathrm{max}}=0.4\ h$Mpc$^{-1}$ we
find $f\sigma_8=0.498$ for z1 NGC and $f\sigma_8=0.497$ for z3 NGC. Now the
bias is larger, and suggests more
significant model misspecification.

\begin{figure}[tbp]
\centering
    \includegraphics[scale=0.28]{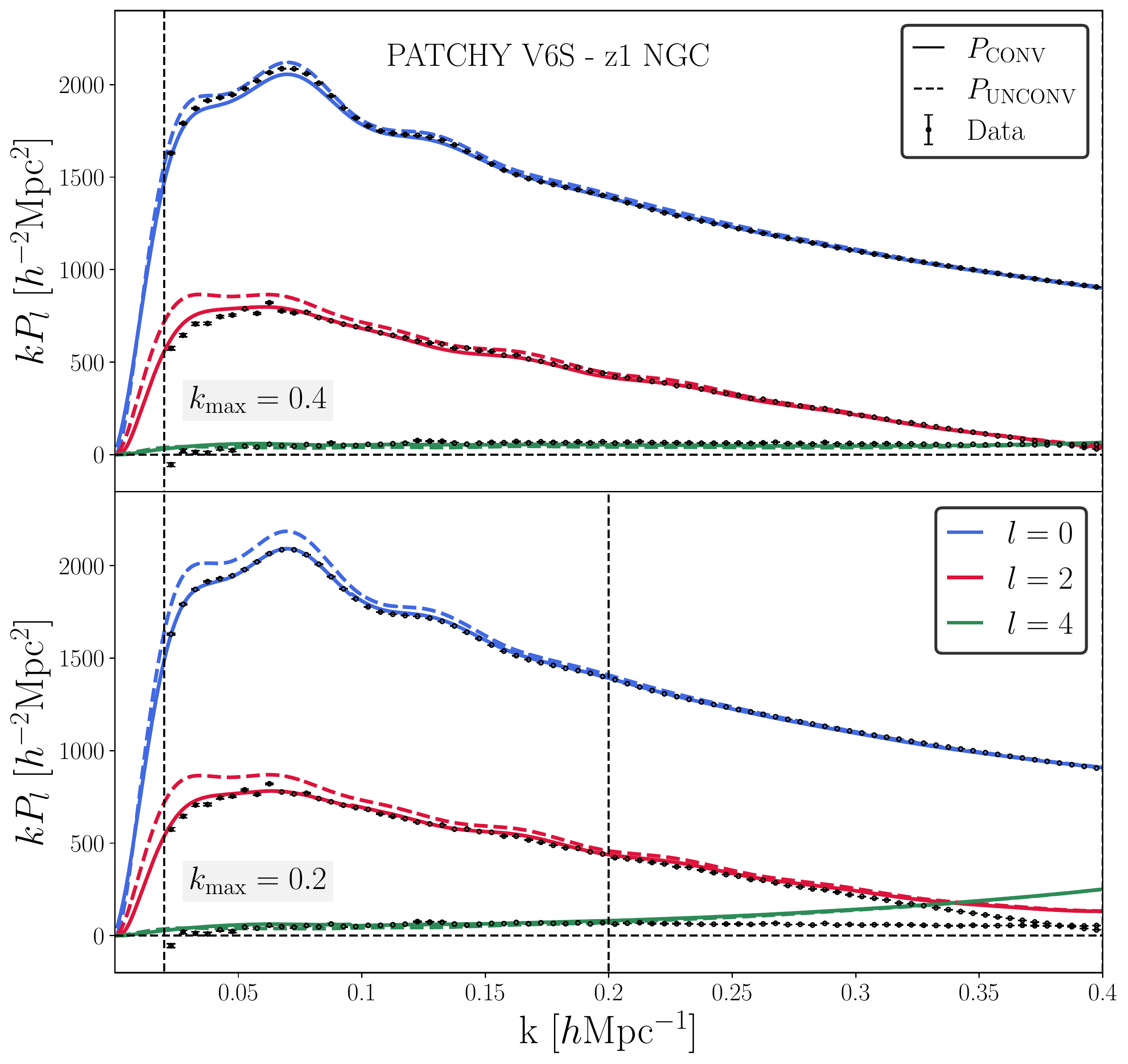}
    \includegraphics[scale=0.28]{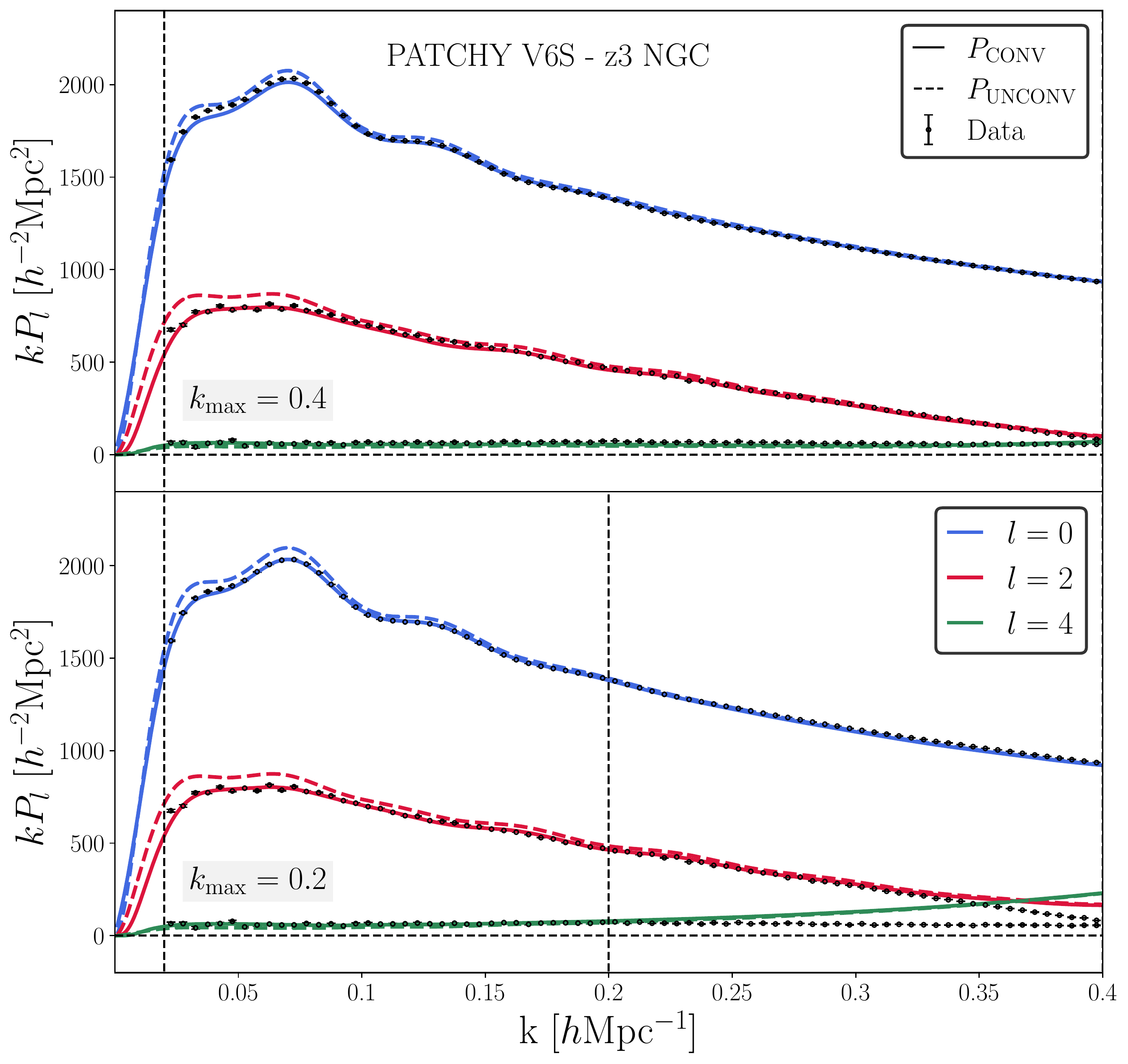}
    \caption{Power spectrum multipole measurements (circular points) of MD-PATCHY z1 NGC and z3 NGC mock catalogues and the best-fit theory models. We take 1000 realizations, and the errors are therefore reduced by $\sqrt{1000}$. Solid and dashed curves indicate convolved and unconvolved best-fit theory curves, respectively. The upper panel figures assume $k_{\mathrm{max}}=0.4\ h$Mpc$^{-1}$, while the lower panel assumes a lower $k_{\mathrm{max}}$ ($0.2\ h$Mpc$^{-1}$ with vertical lines showing the corresponding wavenumber limits to the fits. We find that extending the model to a higher $k_{\mathrm{max}}=0.4\ h$Mpc$^{-1}$ limit makes the model fit worse at low $k$, and as a result an incorrect cosmology may be recovered. We do not observe this issue for $k_{\mathrm{max}}=0.2\ h$Mpc$^{-1}$.}
    \label{fig:kmaxfits}
\end{figure}

To investigate this further,
Figure~\ref{fig:kmaxfits} shows the power spectrum multipole measurements of MD-PATCHY z1 and z3 NGC mock catalogues, along with the best-fit theory lines. Measurements are averaged over 1000 realizations, and the errors are therefore significantly smaller than those of the BOSS survey. Solid and dashed curves indicate convolved and unconvolved best-fit theory curves, respectively. Comparing the upper panel figures (assuming $k_{\mathrm{max}}=0.4\ h$Mpc$^{-1}$) with the lower panel figures ($k_{\mathrm{max}}=0.2\ h$Mpc$^{-1}$), we find that extending the model to a higher $k_{\mathrm{max}}$ limit makes the model fit noticeably worse at low $k$: both monopole and quadrupole fits with lower $k_{\mathrm{max}}$ have better fits for z1 NGC, and similarly monopole fit has with lower $k_{\mathrm{max}}$ has a better fit for z3 NGC.

MD-PATCHY mocks are not based on a real
N-body simulation, and it is unclear if the
galaxy catalogs and the resulting power spectra can be realized in an actual
universe.
For this reason such comparisons against
PATCHY have not
been implemented elsewhere, and it is
unclear whether we should be concerned
given the model is good against real
simulations. On
the other hand, using 1000 mocks enables
one to separate statistical fluctuations from systematics extremely well. In other tests based on one or a few
simulation volumes
the deviations of recovered
cosmological parameters from the truth
were within one statistical
deviation, in which case
it is unclear whether this is a purely
statistical effect that can be
ignored, or it is a sign of model misspecification,
and one must correct for it. As we are
unable to answer whether MD-PATCHY power
spectra
can represent an actual galaxy realization
in a real universe,
we present both results.
However, more caution should be taken when extending to a higher $k_{\mathrm{max}}$, and we argue that a consistent, reliable choice of $k_{\mathrm{max}}$ is one of the major unresolved questions of the RSD analyses. It becomes increasingly difficult to obtain unbiased estimates as we
push to higher $k_{\mathrm{max}}$, simply
because the fits are dominated by the
smallest errors which are close to
$k_{\mathrm{max}}$, but cosmology information is entirely extracted by the
low $k$ asymptote of the fitted model:
even a slight model misspecification at high $k$
can lead to a biased answer at low $k$.
Moreover, since we have to fit more
parameters to high $k$ the choice
of their prior distribution also affects
the fits: even a seemingly innocent
flat prior choices can project onto a
significant bias in the cosmological
parameters.
Different works choose different values, which may partially explain why results from different studies are discrepant in terms of their $f\sigma_8$ measurements.

\section{BOSS DR12 RSD measurements}\label{7}

\begin{figure}[t]
    \includegraphics[scale=0.275]{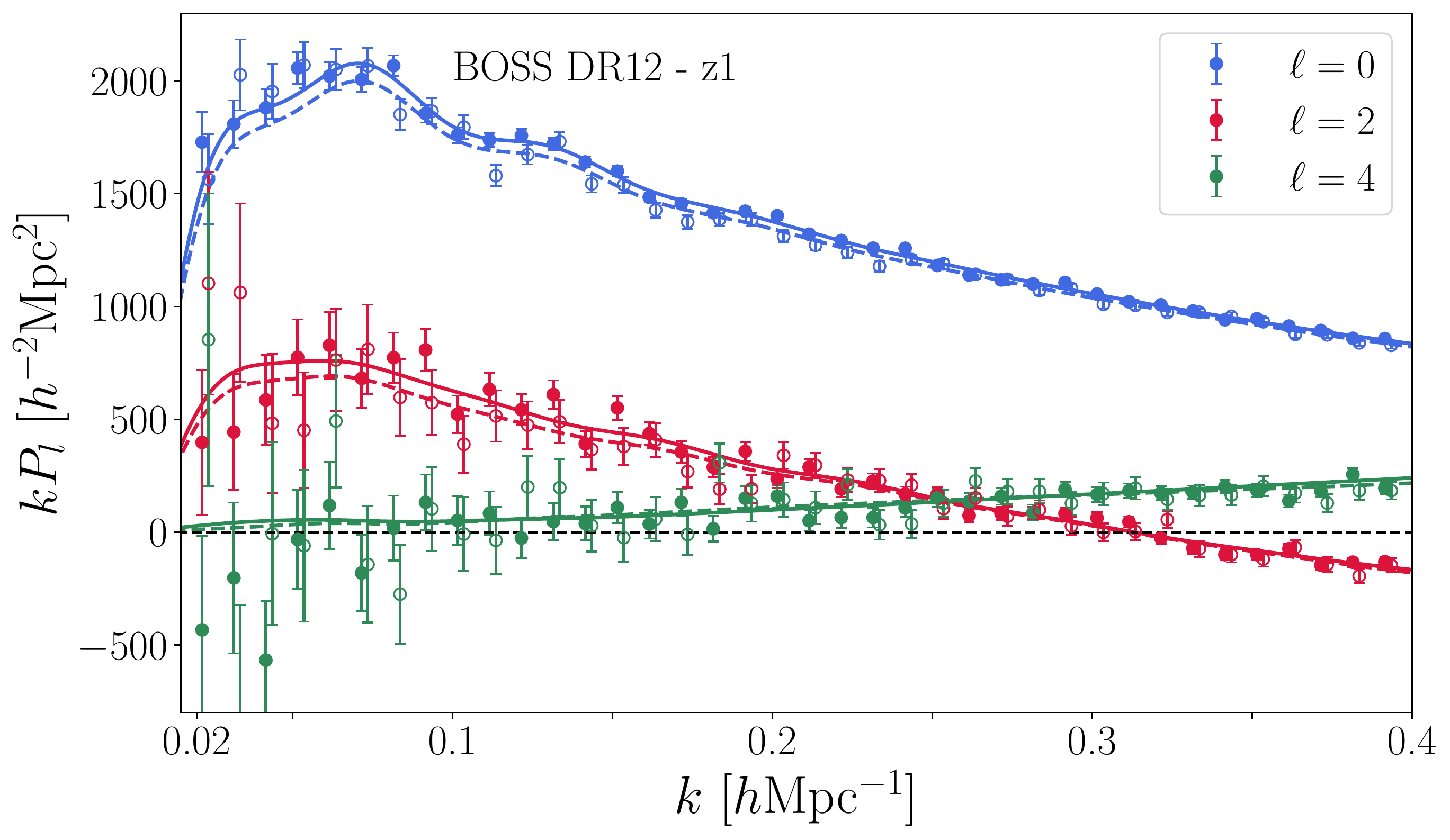}
    \includegraphics[scale=0.275]{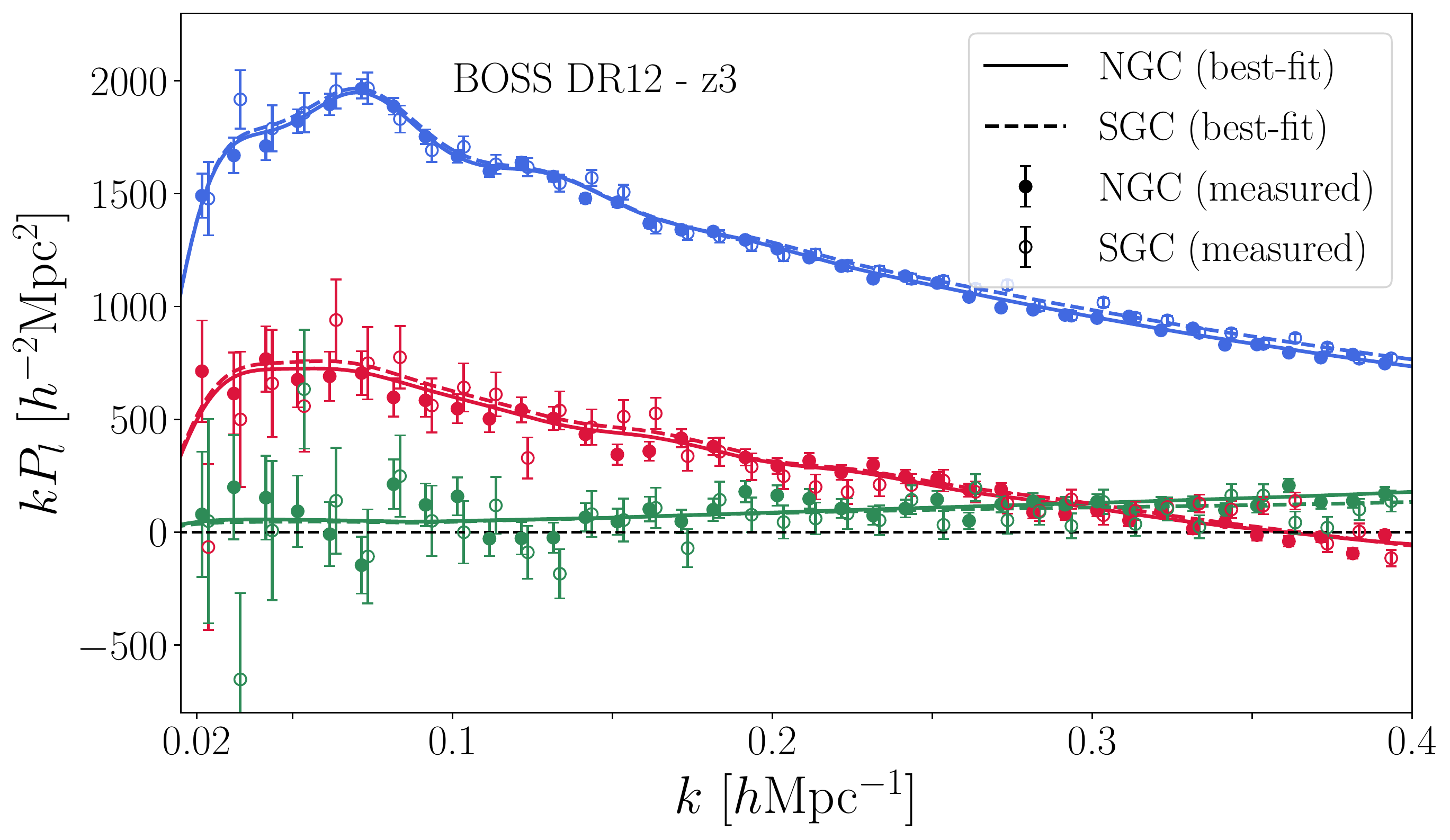}
\caption{The measured galaxy power spectrum multipoles in Fourier space (circular points with error bars) and the best-fit theory curves (solid lines) for BOSS DR12 z1 (\textit{Left} panel) and z3 (\textit{Right} panel) samples. (We only show every other data points for simplicity.) We can fit the model to the monopole (blue), quadrupole (red), and hexadecapole (green), over the wavenumber range $0.02 - 0.4 h$Mpc$^{-1}$, with $\Delta k = 0.005$, but the choice of $k_{\mathrm{max}}$ may affect the cosmological analysis moderately, as discussed in section~\ref{6.1}. Following \cite{Beutler:2021eqq}, we use a consistent definition of the normalization term for both power spectrum and window function.}
    \label{fig:BOSS_poles}
\end{figure}

\begin{figure}[h]
\centering
    \includegraphics[scale=0.28]{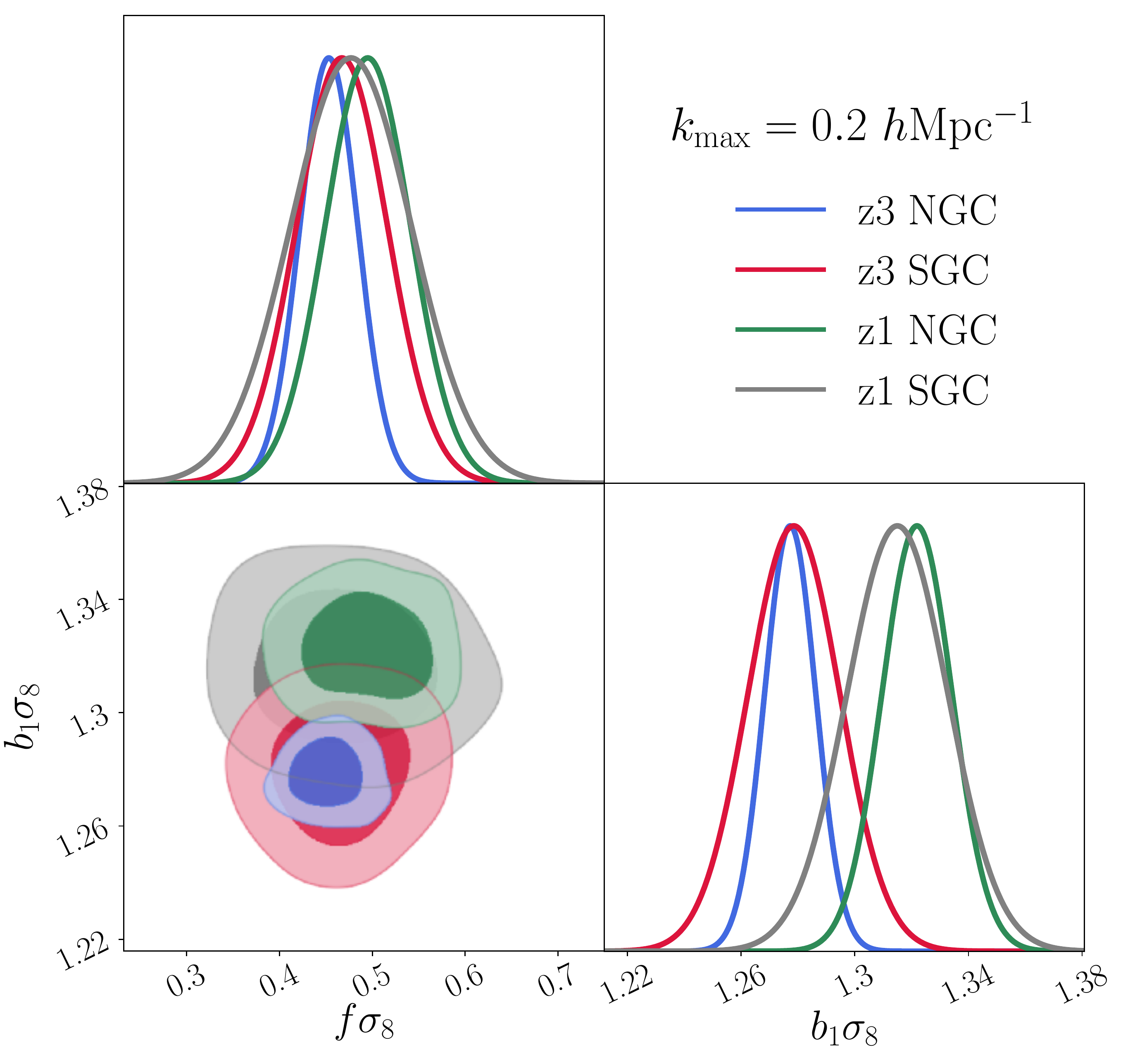}
    \includegraphics[scale=0.3]{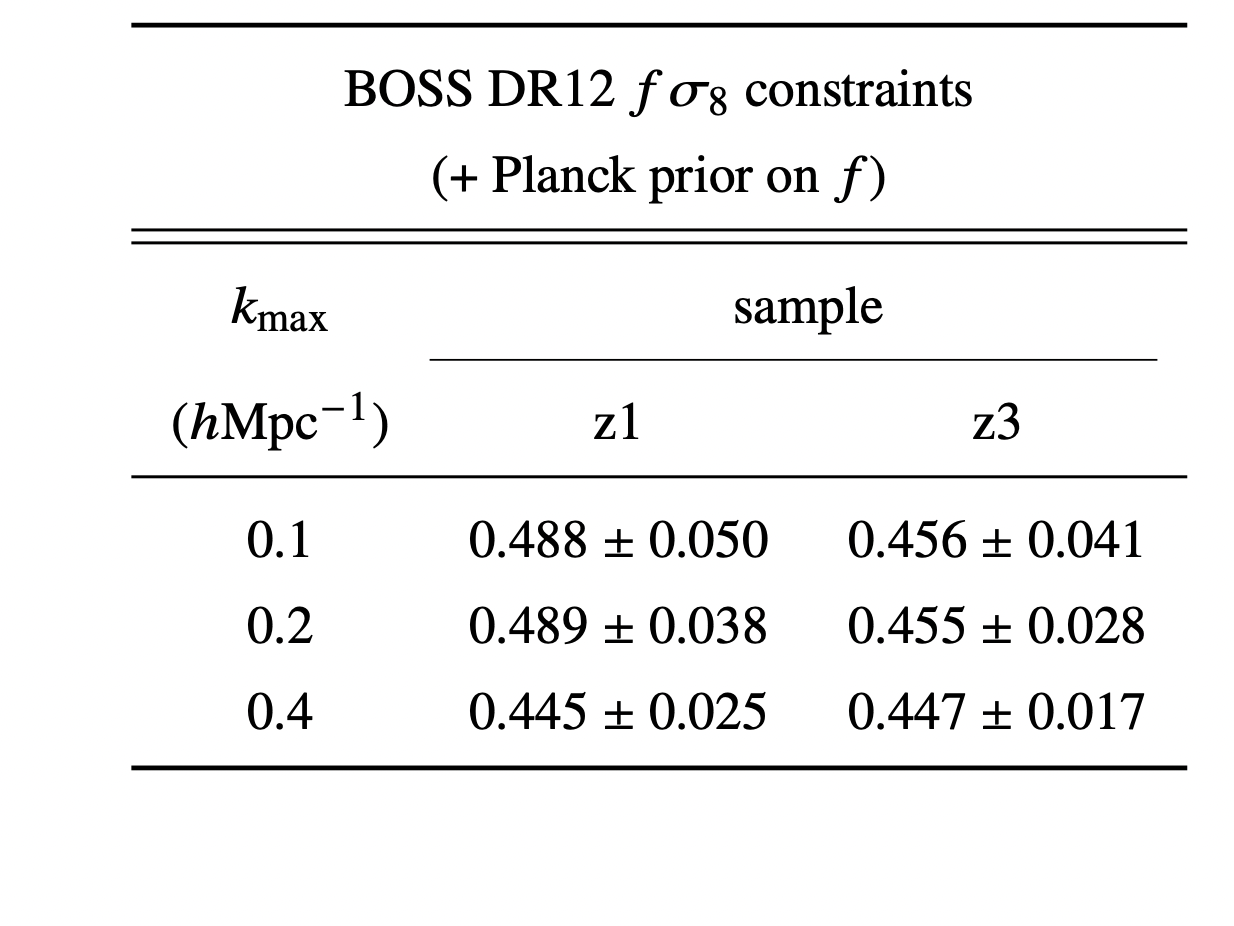}
    \caption{\textit{Left}: 1D and 2D posterior distributions of two selected parameters $f\sigma_8$ and $b_1\sigma_8$ for all galaxy samples with $k_{\mathrm{max}}$=0.2 $h$Mpc$^{-1}$. \textit{Right}: The best-fit $f\sigma_8$ values and their 1$\sigma$ uncertainties for the BOSS DR12 sample in two different redshift bins (including both NGC and SGC sky patches). We put the prior on $f$ using the Planck 2018 prior for $\Omega_m$ \citep{Planck:2018vyg}. For all results, we fit to the monopole, quadrupole, and hexadecapole, assuming the full covariance with the smoothed connected parts, as described in section~\ref{4.3}. We also account for the systematic error resulting from the selection of the underlying template cosmology, as described in section~\ref{3.1}; this increases the overall error by 5\%.}
    \label{fig:BOSS_con}
\end{figure}

In this section, we present the measurements of the BOSS DR12 galaxy power spectrum multipoles in Fourier space. In figure~\ref{fig:BOSS_poles}, we show the measured monopole $P_0(k)$, quadrupole $P_2(k)$, and hexadecapole $P_4(k)$ of z1 and z3 galaxies in both NGC and SGC patches, using the FFT-based galaxy power spectrum estimator described in section~\ref{3.2}. We then fit the RSD model presented in section~\ref{3.1} to the measured multipoles and find that the power spectrum multipoles are accurately modeled, in agreement with~\cite{Hand:2017ilm}. In our fits, we set the minimum wavenumber $k_{\mathrm{min}}$ to 0.02$\ h$Mpc$^{-1}$ for all samples, in order to minimize any large-scale effects of the window function. As described in section~\ref{3.1}, we fix the AP distortion parameters to their fiducial values and constrain 11 model parameters ($f(z_{\mathrm{eff}}), \sigma_8(z_{\mathrm{eff}}), b_{1, c_A}, b_{1, s_A}, b_{1, s_B}, f_s, f_{s_B}, \langle N_{>1,s}\rangle, \sigma_c, \sigma_{s_A}, f^{1h}_{s_B s_B}$), of which two are primarily of our interests: the growth rate $f$ and the amplitude of matter fluctuations $\sigma_8$. The Planck satellite provides a strong prior on $f$ with the tight constraint of the matter density $\Omega_m$ \citep{Planck:2018vyg}, and we therefore put the Planck 2018 prior on $f$ in all subsequent analyses.

\begin{figure}[tbp]
    \includegraphics[scale=0.29]{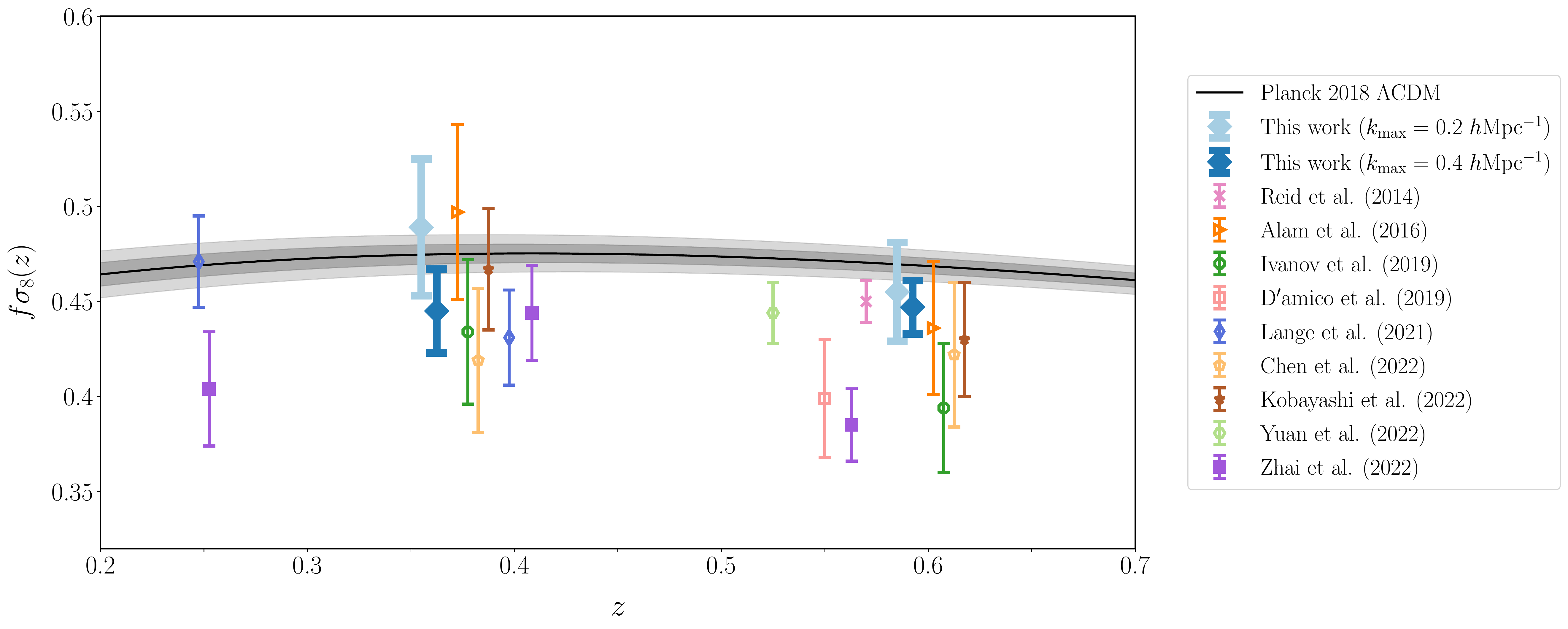}
    \caption{Comparison of $f\sigma_8$ constraints to previous BOSS DR12 measurements \citep{Reid:2014iaa, AlamEtAl17, Ivanov:2019pdj, DAmico:2019fhj, Lange:2021zre, Chen:2021wdi, Kobayashi:2021oud, Yuan:2022jqf, Zhai:2022yyk}, along with the prediction assuming the Planck 2018 $\Lambda$CDM cosmology (black curve with grey shades) \citep{Planck:2018vyg}. We show our main results as thick diamonds points, for two $k_{\mathrm{max}}$ limits: 0.2 (sky blue points) and 0.4 $h$Mpc$^{-1}$ (dark blue points). We show a 
    partial compilation of 
    other results, which used different methodologies, and cannot always be compared in reported errors. Our results, Alam et al. \citep{AlamEtAl17}, Ivanov et al. \citep{Ivanov:2019pdj}, Chen et al. \citep{Chen:2021wdi}, and Kobayashi et al. \citep{Kobayashi:2021oud} present measurements of $f\sigma_8$ for z1 and z3 galaxy samples ($z_{\mathrm{eff}}=0.38$ and $0.61$, respectively), but for graphical purpose they are plotted at different redshifts. Reid et al. \citep{Reid:2014iaa}, D'amico et al. \citep{DAmico:2019fhj}, and Yuan et al. \citep{Yuan:2022jqf} show the constraints on the CMASS sample at $z_{\mathrm{eff}}=0.57, 0.55$ and $0.52$, respectively, Lange et al. \citep{Lange:2021zre} takes the galaxy samples at $z=0.25$ and $0.4$, and Zhai et al. \citep{Zhai:2022yyk} splits the galaxy sample into three redshift bins at $z_{\mathrm{eff}}=0.25, 0.41,$ and $0.56$.}
    \label{fig:comparison}
\end{figure}

Figure~\ref{fig:BOSS_con}, along with the table of $f\sigma_8$ constraints with varying $k_{\mathrm{max}}$ cuts, summarizes the main results of our analysis. Fitting our RSD model to the measured BOSS DR12 multipoles (monopole, quadrupole, and hexadecapole) and marginalizing over all nuisance parameters discussed in section~\ref{3.1}, we obtain the following growth of structure constraints: $f\sigma_8(z_{\mathrm{eff}}=0.38)=0.489 \pm 0.038$ for z1 sample and $f\sigma_8(z_{\mathrm{eff}}=0.61)=0.455 \pm 0.028$ for z3 sample, with $k_{\mathrm{max}} = 0.2\ h$Mpc$^{-1}$. In the table, we also provide the constraints as a function of $k_{\mathrm{max}}$: $f\sigma_8(z_{\mathrm{eff}}=0.38)=0.488 \pm 0.050$ for z1 and $f\sigma_8(z_{\mathrm{eff}}=0.61)=0.456 \pm 0.041$ for z3 with $k_{\mathrm{max}} = 0.1\ h$Mpc$^{-1}$, while $f\sigma_8(z_{\mathrm{eff}}=0.38)=0.445 \pm 0.025$ for z1 and $f\sigma_8(z_{\mathrm{eff}}=0.61)=0.447 \pm 0.017$ for z3 with $k_{\mathrm{max}} = 0.4\ h$Mpc$^{-1}$. For all results, we assume the full covariance with the smoothed connected parts in section~\ref{4.3}, and we also include the systematic error that arises from the selection of the underlying template cosmology, which increases the overall error by 5\%. The left panel of Figure~\ref{fig:BOSS_con} also presents the constraints for $b_1\sigma_8$: with $k_{\mathrm{max}} = 0.2\ h$Mpc$^{-1}$, we obtain $b_1 \sigma_8(z_{\mathrm{eff}}=0.38)=1.320 \pm 0.011$ for z1 and $b_1 \sigma_8(z_{\mathrm{eff}}=0.61)=1.278 \pm 0.009$ for z3 sample.

We can translate our constraints in terms of the parameter $S_8^\gamma \equiv \sigma_8(\Omega_m/0.3)^\gamma$, where $\gamma = \mathrm{dln}f\sigma_8/\mathrm{dln}\Omega_m \simeq 0.78 \cdot (1-\Omega_m(z))$ \cite{Kazantzidis:2018rnb}. For the galaxy samples considered in this work and
for Planck fiducial value of $\Omega_m$ we have $\gamma=0.37$ for z1 and 0.28 for z3, respectively. We find $S_8^\gamma = 0.821 \pm 0.040$ and $0.824 \pm 0.056$ with $k_{\mathrm{max}} = 0.2$ and $0.1\ h$Mpc$^{-1}$, respectively, consistent with Planck's value $S_8^\gamma \sim 0.83 \pm 0.01$ \citep{Planck:2018vyg}. Extending to a higher $k_{\mathrm{max}}$ of $0.4\ h$Mpc$^{-1}$, we get $S_8^\gamma = 0.786 \pm 0.025$, about 1.8$\sigma$ lower than the Planck constraint. In section~\ref{6.1} we argued that extending the model to $k_{\mathrm{max}}$ of $0.4\ h$Mpc$^{-1}$ may lead to model misspecification. We note that 
for RSD $\gamma$ is lower than in weak lensing (where typically $\gamma>0.5$), so 
RSD and weak lensing do not 
measure the same combination of $\sigma_8$ and $\Omega_m$ as encoded in $S_8^\gamma$. 
Specifically, RSD is more 
sensitive to $\sigma_8$ only, 
specially at higher 
redshifts. 

Figure~\ref{fig:comparison} compares our $f\sigma_8$ measurements to previous BOSS DR12 results in the literature \citep{Reid:2014iaa, AlamEtAl17, Ivanov:2019pdj, DAmico:2019fhj, Lange:2021zre, Chen:2021wdi, Kobayashi:2021oud, Yuan:2022jqf, Zhai:2022yyk}, along with the constraint assuming the Planck 2018 $\Lambda$CDM cosmology \citep{Planck:2018vyg}. \citep{AlamEtAl17} divides the BOSS galaxies into three redshift bins ($z_1, z_2$ and $z_3$) and provides the ``consensus'' $f\sigma_8$ constraints by combining measurements from seven companion papers. Our constraints are not in tension with the consensus analysis (with $k_{\mathrm{max}} = 0.2\ h$Mpc$^{-1}$, $0.21\sigma$ lower for z1 and $0.68\sigma$ higher for z3) or with the Planck 2018 predictions (with $k_{\mathrm{max}} = 0.2\ h$Mpc$^{-1}$, $0.37\sigma$ higher for z1 and $0.49\sigma$ lower for z3), while providing one of the tightest constraints on $f\sigma_8$
among recent works.

\begin{figure}[t]
    \centering
    \includegraphics[scale=0.33]{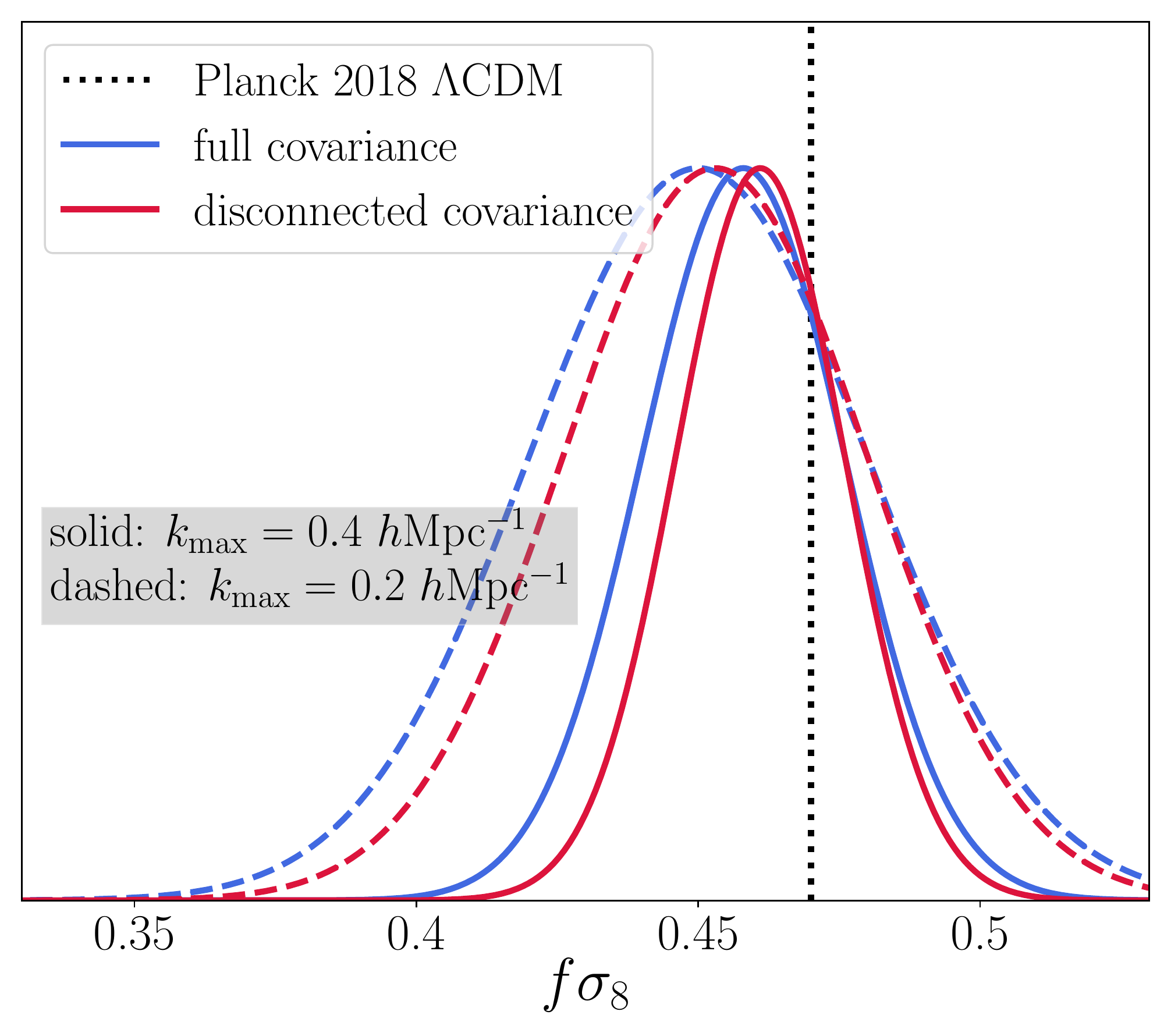}
    \includegraphics[scale=0.33]{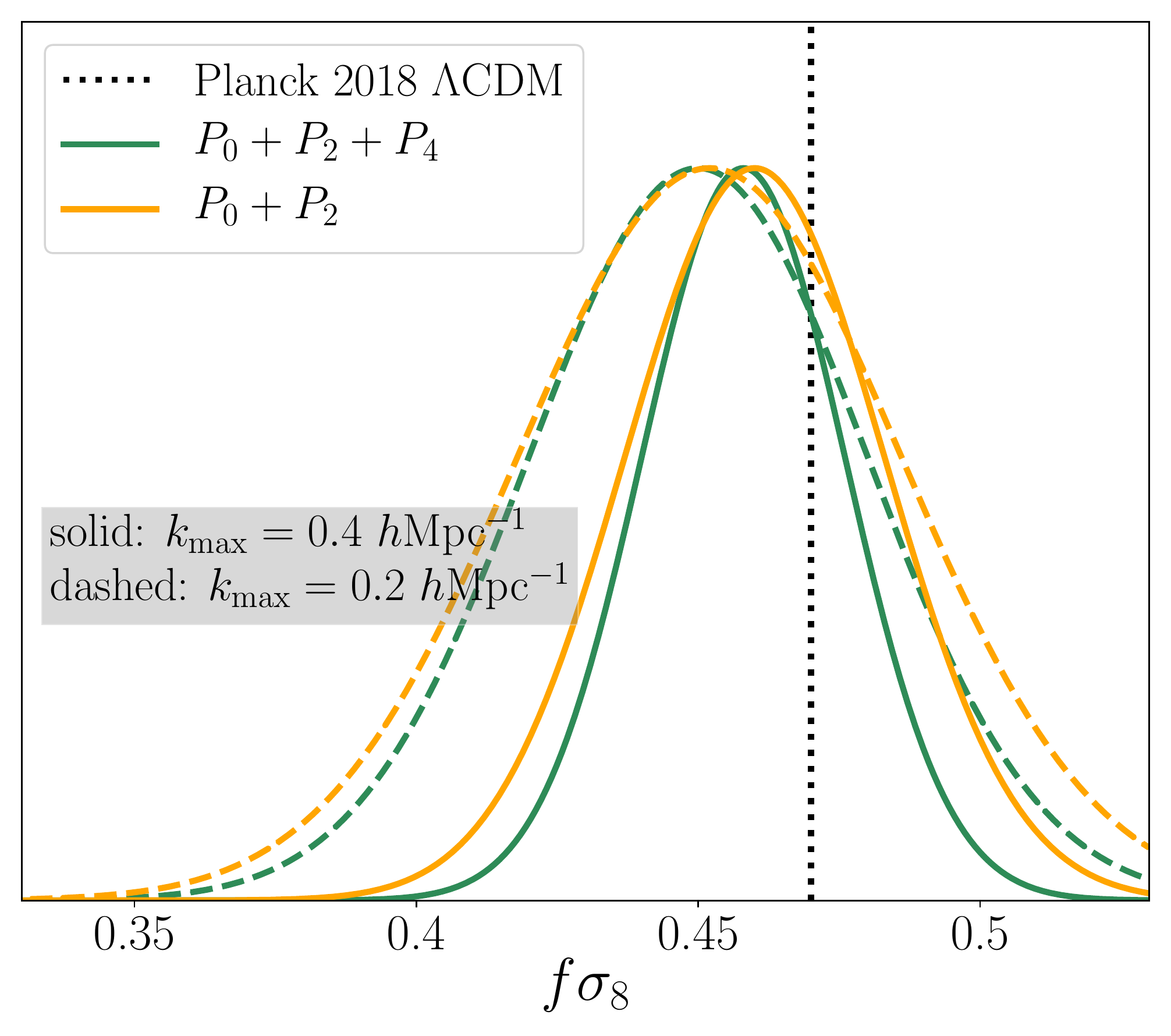}
    \caption{$f\sigma_8$ constraints of the BOSS DR12 z3 NGC sample, varying dataset, wavenumber range, and covariance matrices. \textit{Left}: Constraints obtained with different covariances matrices in section~\ref{sec4}. Adding the connected parts (blue) to the analytic disconnected covariance (red) weakens our constraints by 10-20\%. We also show the impact of including wavenumbers in a wider range. Dashed curves assume $k_{\mathrm{max}}=0.2$ $h$Mpc$^{-1}$, and extending it to 0.4 $h$Mpc$^{-1}$ (solid curves) improves our constraints considerably.
    \textit{Right}: Results obtained with the inclusion (green) or exclusion (orange) of the hexadecapole to quantify its impact on $f\sigma_8$ constraints. Excluding the hexadecapole inflates its standard deviation by 15-30\%. For both results, we assume the full covariance matrix with the connected part.}
    \label{fig:BOSS_con2}
\end{figure}

Figure~\ref{fig:BOSS_con2} shows how different combination of dataset, wavenumber range, and covariance matrices may affect the cosmological analysis. With $k_{\mathrm{max}} = 0.2\ h$Mpc$^{-1}$, we obtain $f\sigma_8 = 0.450 \pm 0.032$ for z3 NGC, assuming the full covariance including connected parts. Extending $k_{\mathrm{max}}$ to $0.4\ h$Mpc$^{-1}$ tightens the constraint significantly, while shifting the best-fitting parameter mean modestly ($f\sigma_8 = 0.458 \pm 0.021$). The dotted curves in the left panel present the constraint assuming the analytic disconnected covariance ($f\sigma_8 = 0.453 \pm 0.029$ and $0.461 \pm 0.018$ for $k_{\mathrm{max}}=0.2$ and $0.4\ h$Mpc$^{-1}$, respectively), and we thus find that including the connected parts inflates its standard deviation by 10-20\%.

In the right panel of Figure~\ref{fig:BOSS_con2}, we show results of fitting only the monopole $P_0$ and quadrupole $P_2$ to quantify the impact of including hexadecapole $P_4$ on our constraints on the growth of structure. With $P_0$ and $P_2$ only, we obtain $f\sigma_8 = 0.452 \pm 0.035$ and $0.460 \pm 0.025$ for $k_{\mathrm{max}}=0.2$ and $0.4\ h$Mpc$^{-1}$, respectively, and find that the best-fit parameter mean remains consistent, while the standard deviation of $f\sigma_8$ increases by 15-30\%, which is consistent with \cite{BeutlerEtAl17a}. Therefore, we include the hexadecapole $P_4(k)$ in our main analysis (presented in Figure~\ref{fig:BOSS_con}) because this improves RSD constraints significantly, as reported earlier in \mbox{\cite{Beutler:2016arn}, \cite{Grieb2016},} and \cite{Hand:2017ilm}.

\begin{figure}[t]
\includegraphics[scale=0.325]{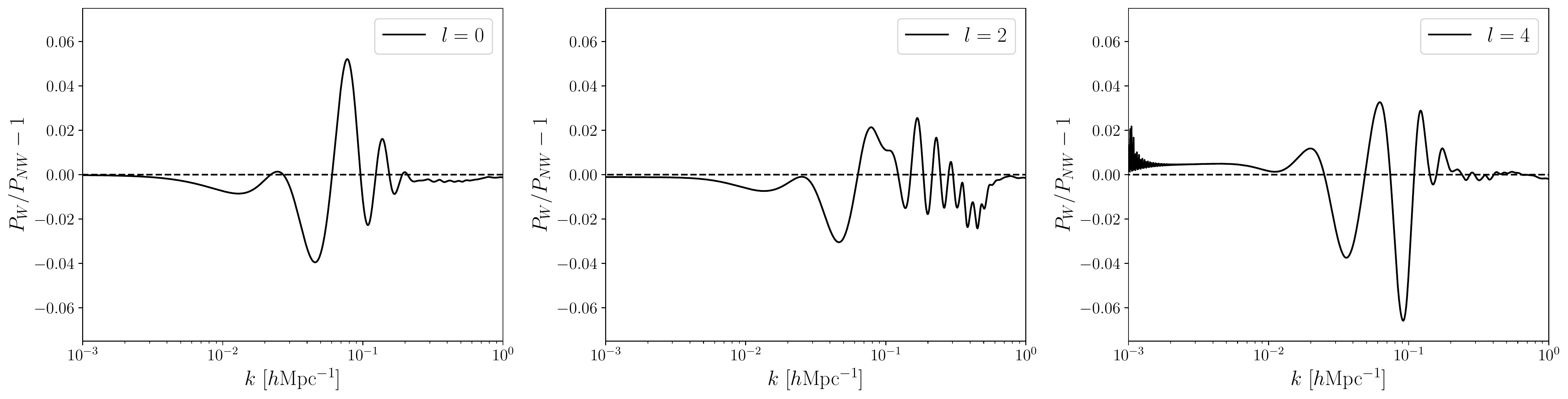}
\caption{The ratio of the wiggle ($P_{\mathrm{W}}$) to the no-wiggle ($P_{\mathrm{NW}}$) poles of the BOSS DR12 z3 NGC galaxy power spectrum using the best-fit cosmology. The ratio for the monopole, quadrupole, and hexadecapoles are shown in the left, middle, and right panels, respectively.}
\label{fig:BAOdamping}
\end{figure}

\begin{figure}[h]
    \includegraphics[scale=0.3]{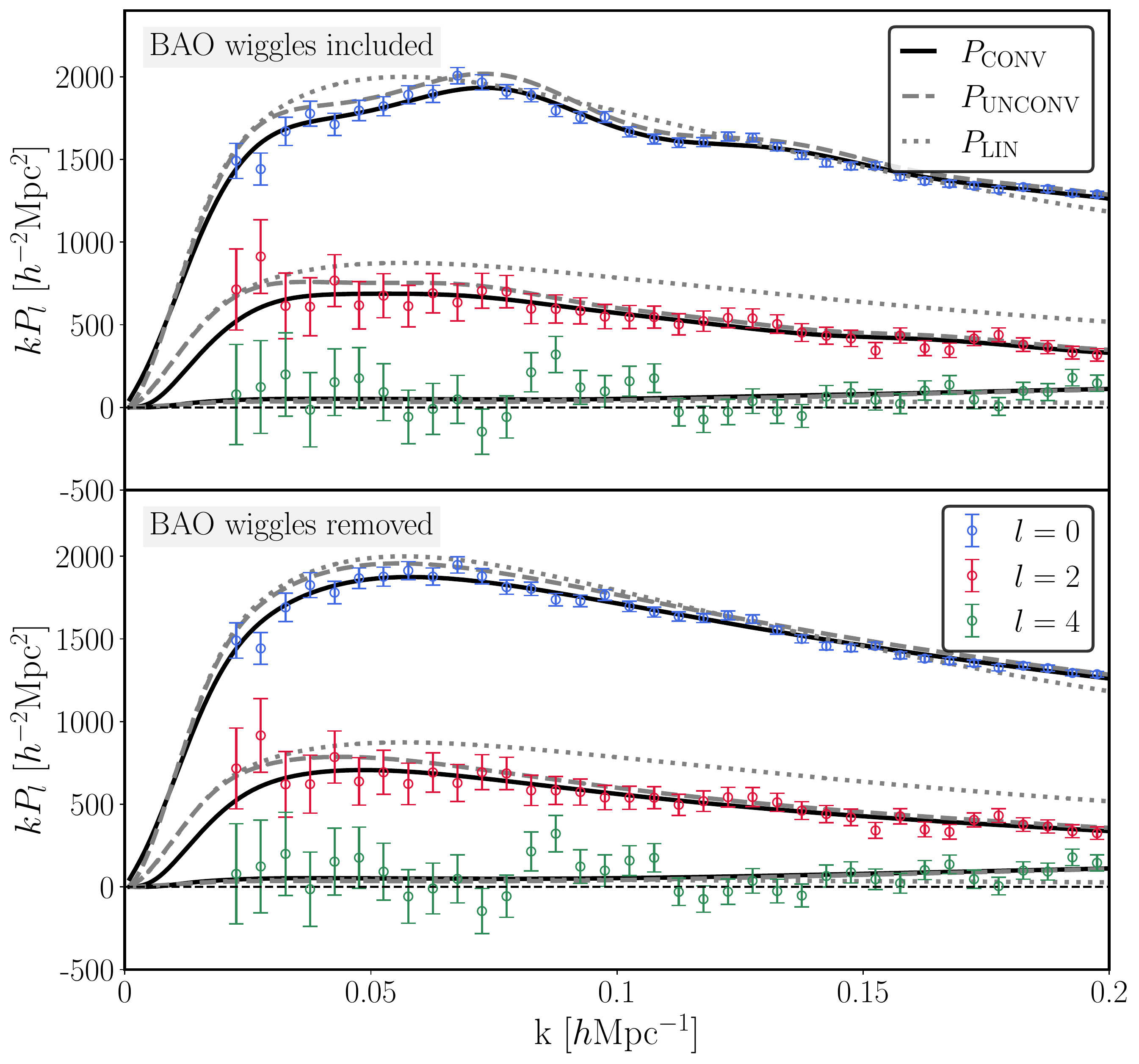}
    \includegraphics[scale=0.22]{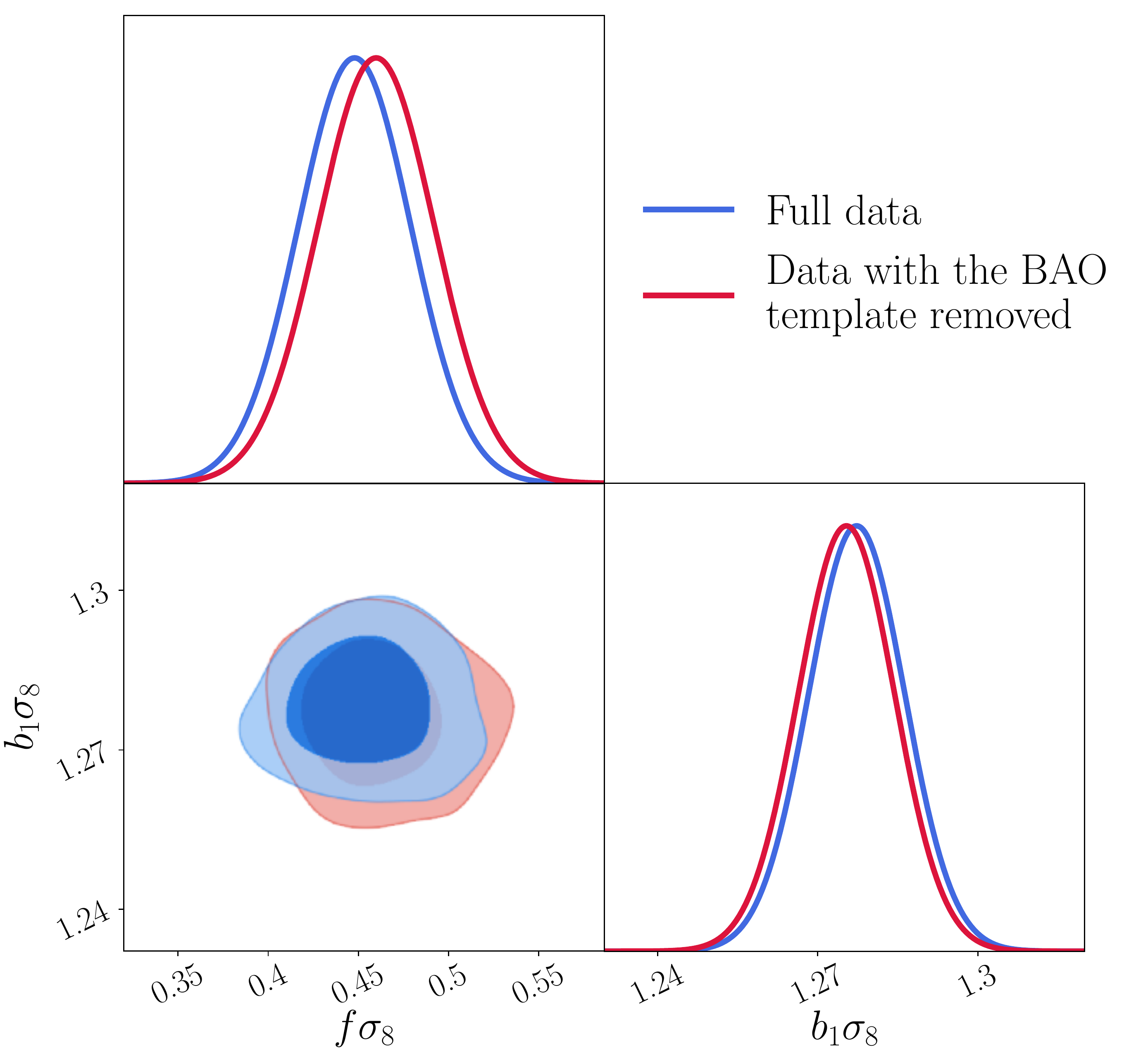}
    \caption{\textit{Left}: (Top panel) Power spectrum multipole measurements of the BOSS DR12 z3 NGC sample with BAO wiggles. Solid curves indicate the best-fit theory model, convolved with the window function ($P_{\mathrm{CONV}}$). Dashed and dotted curves indicate best-fit unconvolved ($P_{\mathrm{UNCONV}}$) and linear ($P_{\mathrm{LIN}}$) theory models, respectively. (Bottom panel) We take the BAO wiggle template from figure~\ref{fig:BAOdamping} and subtract it from the measured data so that we can get the ``no-wiggle'' measurements. All best-fit theory lines also assume the models without BAO wiggles.  \textit{Right}: The model constraint with the full power spectrum and the no-wiggle fit with the fixed fiducial BAO template. Fitting to $k_{\mathrm{max}} = 0.2 h$Mpc$^{-1}$, we obtain $f\sigma_8(z_{\mathrm{eff}}=0.61)=0.450 \pm 0.032$ and $0.459 \pm 0.033$ with and without BAO, respectively. Therefore, we argue that fixing the BAO information with the fiducial BAO template does not affect the cosmology considerably, which eliminates the need to do a joint power spectrum BAO
    covariance analysis on mocks.}
    \label{fig:BAOconst}
\end{figure}

In Figure~\ref{fig:BAOdamping} and Figure~\ref{fig:BAOconst}, we study the sensitivity of cosmological analysis to adding a fixed BAO template to the galaxy power spectrum model. The motivation for this analysis is that  there is BAO information in 
reconstruction, but there is also 
BAO information in P(k). The two 
are not independent, and BAO information should 
not be counted twice. In the past SDSS team used covariance matrix
from mocks to identify the amount 
of correlation. While we do not do BAO 
analysis here since we fix cosmology 
parameters other than $f\sigma_8$, it is 
worth asking how to avoid double counting 
of BAO information in such analysis 
other than building the 
full covariance matrix. 

Figure~\ref{fig:BAOdamping} shows the ratio of the wiggle ($P_{\mathrm{W}}$) to the no-wiggle ($P_{\mathrm{NW}}$) poles of the BOSS DR12 z3 NGC galaxy power spectrum using the best-fit cosmology; we use the wiggle only nonlinear damping from \cite{2016Vlah} for BAO models. With this ratio, we create a fixed template for BAO wiggles and subtract it from the multipole measurements to get no-wiggle data. The left panel of Figure~\ref{fig:BAOconst} shows z3 NGC multipole measurements (circular points with error bars): full data with BAO wiggles included (top) and data with BAO wiggles removed (bottom). Subsequently, we fit the theory model to both datasets, full and no-wiggle, and compare their $f\sigma_8$ constraints. The right panel of Figure~\ref{fig:BAOconst} shows that the model constraints for both samples. Fitting to $k_{\mathrm{max}} = 0.2 h$Mpc$^{-1}$, we obtain $f\sigma_8(z_{\mathrm{eff}}=0.61)=0.450 \pm 0.032$ and $0.459 \pm 0.033$ with and without BAO, respectively; we find that removing BAO information does not shift the model parameter constraints significantly and fixing the BAO information with the fiducial template only minimally affects the cosmology. This result suggests a simplified large-scale structure
analyis where the BAO information is
completely independent of the
de-wiggled power spectrum analysis,
without a need to compute their
covariance from mocks, as is commonly done when 
combining BAO reconstruction and RSD analyses.

\section{Conclusion}\label{8}

We have presented the analysis of the SDSS-III BOSS DR12 galaxy sample, employing the redshift-space galaxy power spectrum model of \cite{Hand:2017ilm}, which can accurately model the monopole, quadrupole, and hexadecapole down to small scales. With the Planck 2018 prior on the growth rate $f$, we obtain 7.7\% and 6.2\% constraints on $f\sigma_8$ for BOSS DR12 low-redshift z1 ($z_{\mathrm{eff}}=0.38$) and high-redshift z3 ($z_{\mathrm{eff}}=0.61$) samples, respectively, fitting to $k = 0.2\ h$Mpc$^{-1}$. The combined error is 5\%. Extending the wavenumber range to $k_{\mathrm{max}} = 0.4\ h$Mpc$^{-1}$, we find significant improvement in our constraint: 5.6\% and 3.8\% constraints on $f\sigma_8$ for z1 and z3 samples, respectively, which correspond to an overall 3.2\% constraint. However, tests on MD-PATCHY mock catalogues suggest that the model fit to $k_{\mathrm{max}} > 0.2\ h$Mpc$^{-1}$ may not be reliable, and more caution should be taken when extending to smaller scales. This is further supported
by strong running of $f\sigma_8$ with
$k_{\mathrm{max}}$ for $z1$
from
0.2 to 0.4 $h$Mpc$^{-1}$ seen in Table \ref{fig:BOSS_con}, in contrast to
little or no running from 0.1 to 0.2 $h$Mpc$^{-1}$.
We argue that a consistent choice of $k_{\mathrm{max}}$ is the main challenge of RSD analyses. With respect to the Planck 2018 $\Lambda$CDM cosmology predictions or the DR12 final consensus results, we find no tension in our $f\sigma_8$ constraints for $k_{\mathrm{max}} = 0.2\ h$Mpc$^{-1}$,
while being 2 sigma lower for
$k_{\mathrm{max}} = 0.4\ h$Mpc$^{-1}$.

Figure \ref{fig:comparison} presents a review of literature.
Most of the measurements are below Planck,
and this has been interpreted as evidence
of $\sigma_8$ tension with Planck from
BOSS RSD (e.g, \cite{Ivanov:2021zmi, 2022PhRvD.105d3517P, zhang2022boss, chen2022cosmological}). However, all the analyses are
based on the same underlying data (galaxy
positions and redshifts), and one cannot
simply average these different analyses.
The spread
of the results is indicative of
either model misspecification of some or all of the
models, of the influence
of the choice of priors, or of the
choices made in the analysis, such as
the scale cut $k_{\mathrm{max}}$ in power spectrum analysis, power spectrum versus correlation function versus wedges analysis, as well as several
additional choices.
It
is perhaps disturbing that the
spread is so large given that
the underlying data are the same: at
one end of the spectrum our
results at $k_{\mathrm{max}}=0.2\ h$Mpc$^{-1}$ 
agree with Planck within 0.3 sigma.
At the other end of the spectrum some
analyses disagree with Planck at 3-4
sigma. These differences need to
be understood so that we can establish
the existence or absence of $\sigma_8$ tension in cosmology. In this paper we argued that 
simplifying the analysis such that only 
one parameter is being fitted to RSD 
provides a way to do the comparison 
against Planck that is less influenced 
by prior volume and model misspecification. 
Furthermore, we argue that AP parameter
is strongly constrained by BAO reconstrution and 
Planck, and should not be allowed to be varied
freely in RSD analysis. 

We developed the hybrid covariance matrix which combines the analytic disconnected (or more conventionally, ``Gaussian'') part \citep{Li2019}, which accounts for the window function effect, and simulation-based connected part, smoothed by including up to four principal components. We demonstrate that the difference between the mock covariance and the hybrid covariance is clean, and the hybrid covariance is free from noise and hence more appropriate to be used in the likelihood analysis. Additionally, we show that using the disconnected covariance matrix underestimates the cosmological parameter constraints by 10-20\%.

Furthermore, we provide growth of structure constraints without BAO information by constructing a fixed template for BAO wiggles and subtracting it from the BOSS DR12 multipole measurements. Comparing the constraints with and without BAO wiggles, we conclude that removing BAO information does not noticeably shift the cosmological parameter constraints and hence fixing BAO information with the fiducial template affects the cosmology only minimally.
This method enables combining the
power spectrum and BAO likelihoods
independently, rather than computing
their covariance matrix with mock
simulations.

Finally, we note that the galaxy power spectrum model and analysis pipeline used in this work can be useful for extracting cosmological information from the next-generation galaxy redshift surveys, such as the Dark Energy Spectroscopic Instrument (DESI) \cite{DESI2016} and the Euclid \cite{2011arXiv1110.3193L} and Roman \cite{WFIRST2018} missions.  DESI Emission Line Galaxy sample is expected to provide considerably larger constraining power, as its bias is lower and satellite fraction higher than the BOSS high-redshift z3 sample \cite{DESI2016}.

\acknowledgments

We thank Martin White, Stephen Chen, and Joe DeRose for useful comments on the manuscript. US is supported by the National Science Foundation under Grant Numbers 1814370 and NSF 1839217, and by NASA under Grant Number 80NSSC18K1274.
The Flatiron Institute is supported by the Simons Foundation.

\bibliographystyle{JHEP}
\bibliography{main}

\end{document}